\documentclass{elsarticle}

\usepackage{natbib} 
\usepackage{verbatim}
\usepackage{graphicx} 
\usepackage{array} 
\usepackage{paralist} 
\usepackage{verbatim} 
\usepackage{pdflscape} 
\usepackage{amsmath} 
\usepackage{float} 
\usepackage{algorithm}
\usepackage{algorithmic}
\usepackage{amssymb}
\usepackage{epstopdf}
\usepackage{pdflscape}
\usepackage{algorithmic}
\usepackage{verbatim}
\usepackage{enumerate}
\usepackage{arydshln}
\usepackage[T1]{fontenc}
\usepackage[utf8]{inputenc}
\usepackage{geometry}
\usepackage{caption}
\usepackage{subcaption}

\begin{document}
\begin{frontmatter}

\title{Low-Order Mathematical Modelling of Electric Double Layer Supercapacitors Using Spectral Methods}

\author[a]{Ross Drummond }
\author[a]{David A. Howey}
\author[a]{Stephen R. Duncan \corref{mycorrespondingauthor}}

\cortext[mycorrespondingauthor]{Corresponding author. Tel +44 1865 283261. Fax 1865 273906}
\ead{stephen.duncan@eng.ox.ac.uk}

\address[a]{Department of Engineering Science, University of Oxford, Oxford, UK, OX1 3PJ}

\begin{abstract}
This work investigates two physics-based models that simulate the non-linear partial differential algebraic equations describing an electric double layer supercapacitor. 
In one model the linear dependence between electrolyte concentration and conductivity is accounted for, while in the other model it is not.
A spectral element method is used to discretise the model equations and it is found that the error convergence rate with respect to the number of elements is faster compared to a finite difference method.
The increased accuracy of the spectral element approach means that, for a similar level of solution accuracy, the model simulation computing time is approximately 50\% of that of the finite difference method. 
This suggests that the spectral element model could be used for control and state estimation purposes.
For a typical supercapacitor charging profile, the numerical solutions from both models closely match experimental voltage and current data. 
However, when the electrolyte is dilute or where there is a long charging time, a noticeable difference between the numerical solutions of the two models is observed. 
Electrical impedance spectroscopy simulations show that the capacitance of the two models rapidly decreases when the frequency of the perturbation current exceeds an upper threshold.
\end{abstract}


\end{frontmatter}

\section{Introduction}

This paper develops a new spectral element implementation of two non-linear models that describe the behaviour of an electric double layer supercapacitor
Supercapacitors are electrical energy storage devices for high-power applications \cite{sharma2010review,burke2000ultracapacitors}. In contrast to conventional dielectric capacitors, supercapacitors store their energy using the electric double layer (EDL) phenomenon with high specific surface area electrodes. Storing energy in this manner increases the energy density, while still retaining the inherently high power density characteristic of capacitors. 

Supercapacitors have been successfully implemented in a range of applications including grid stabilisation \cite{srithorn2008power} and hybrid electric vehicle power systems \cite{Wu20147885}.
The growing popularity of supercapacitors has necessitated a demand for new models capable of capturing their dynamics accurately. 
Such models are useful for design predictions, online estimation and control. In the literature, several models have already been proposed, with these models being generalised into the two types, equivalent circuit and physics based. 

Equivalent circuit (EC) models such as \cite{zubieta2000characterization} and \cite{rafik2007frequency} use a parameterised resistor-capacitor (RC) circuit to represent the electrical behaviour of the supercapacitor. The main advantage of this approach is that the resulting model is simple, making ECs a popular modelling approach. However, representing the complex dynamics of an electrochemical device by a RC circuit can have limitations. Firstly, the states of the model have no direct physical meaning, making it difficult to infer any understanding of the device from the model. By treating the supercapacitor as a black box in this manner, developing effective control systems becomes problematic and improving the model becomes more challenging. Secondly, the fact that EC models are based upon a parameterisation of a RC circuit means that they are only applicable to one operating condition and any deviation reduces the applicability of the model.

Physics based models instead use a set of conservation and diffusion equations to describe the dynamics of the system. This approach generally involves using a numerical method to solve a system of partial differential equations (PDEs) coupled with algebraic constraints that describe the diffusion and conservation of ions in the supercapacitor. 
Such models are more generally applicable than the equivalent circuit approach, making model tuning, control and development more intuitive. However, the model is based on a set of PDEs, whose solution is much more complicated to compute than the ordinary differential equations (ODEs) of the EC approach. Subsequently, physics-based models are both more complex and computationally burdensome, a problem which has hindered their adoption.

Examples of such physics-based supercapacitor models include \cite{verbrugge2005microstructural} which compares model numerical solutions, obtained using the code of \cite{verbrugge1994finite}, to experimental data.
The multi-physics software COMSOL was used in \cite{madabattulamodeling} for a supercapacitor model with non-binary electrolyte. A single-domain, volume-averaging approach using finite elements was implemented in \cite{allu2014generalized}, whose solution could be extended to higher spatial dimensions. 
In \cite{romero2010reduced} a comparison of the performances of finite difference, finite element as well spectral methods for a linearised version of the supercapacitor model of \cite{verbrugge2005microstructural} was carried out. The spectral methods were found to perform best in this application, being the most accurate for a given number of elements.
In addition to numerical methods, an analytical solution for the supercapacitor PDEs, limited to the constant current and impedance spectroscopy operating conditions, is given in \cite{srinivasan1999mathematical}. A review of available commercial software for modelling supercapacitors can be found in \cite{johansson2008comparison}.

In the related field of lithium ion battery modelling, there has already been a substantial amount of work on numerical techniques for solving physics based models. In particular, the use of spectral methods has been demonstrated by various authors \cite{cai2012lithium, northrop2011coordinate} and \cite{bizeray2013advanced}. In \cite{cai2012lithium}, spectral methods were applied across individual finite elements and in \cite{northrop2011coordinate} a pseudo-spectral method with Jacobi polynomial basis functions was used, while \cite{bizeray2013advanced} used a unified approach involving Chebyshev polynomials, with the same method being used to solve all of the equations. In these papers, spectral methods were found to give a marked reduction in model complexity without loss in accuracy when compared to a benchmark finite difference method as well as the COMSOL finite element solver.

As well as being used to capture the dynamic response of a supercapacitor, another important use of the models is to be incorporated within an observer to increase the accuracy of state estimation. 
 An Extended Kalman Filter was applied to an EC model of a supercapacitor in \cite{nadeaustate}, improving the energy prediction when compared to the straightforward $E = \frac{1}{2}CV^2$ approach, where $E$ is stored energy, $C$ is capacitance and $V$ is voltage.
In the related field of lithium ion batteries there has been extensive study on observer design for power management systems \cite{chaturvedi2010algorithms}. In \cite{suthar2013optimal}, an online implementation of a non-linear moving horizon estimator for a reduced order battery model is presented and then used to solve the optimal control problem of maximising the amount of charge stored in a given amount of time.

In this paper, two non-linear models are investigated that simulate the differential algebraic equations that describe a supercapacitor. 
The first model has a logarithmic non-linearity due to the Nernst-Plank relation \cite{verbrugge2005microstructural} while the second has a coupled quadratic (state) non-linearity that accounts for the linear dependence between electrolyte conductivity and concentration.
The accuracy of the spectral element and finite difference numerical discretisation methods applied to these models is also investigated, with a finite element solution from COMSOL containing a large number of elements being used as a reference solution.
The focus of the models was to be of low order whilst retaining the physical non-linearity as much as possible, so as to give an improved mathematical description of the supercapacitor. A low-order model has the advantage of being less computationally burdensome, making a real-time implementation possible, and also reduces the number of states needed to be estimated by an observer. 

The paper is structured as followed. In Section 2, the governing equations of the supercapacitor models are introduced and the various assumptions and mathematical details are explained. Spectral methods are introduced in Section 3 with a brief summary of their convergence and stability properties. Finally, numerical simulation results of the various models are presented in Section 4.

\section{Mathematical Description of Supercapacitor Model}

The supercapacitor models considered in this paper are based on the four coupled partial differential algebraic equations given in \cite{verbrugge2005microstructural}.
The first of these is the Nernst-Plank equation \cite{kilic2007steric}, describing the diffusion and migration of ions in the liquid electrolyte phase
\begin{equation}\label{eq:NernstPlank}
U_j = -D_j \Bigg(\frac{\partial c_j}{\partial x}-\zeta_j \frac{F}{RT} c_j \frac{\partial \Phi_2}{\partial x}\Bigg)
\end{equation}
with the subscript $j = {1,2}$ referring to the positive and negative ions respectively. This equation can be re-written in terms of the variables $c,\Phi_1, \Phi_2$ and $i_2$ as
\begin{equation}\label{eq:1}
i_2 = -\kappa \frac{\partial \Phi_2}{\partial x} - \kappa \Bigg( \frac{t_+-t_-}{f} \Bigg)\frac{\partial \text{ ln }c}{\partial x}
\end{equation}
using the relations
\begin{subequations}
\begin{equation}\label{eq:IonicFlux}
U_j = \frac{i_2}{\zeta_jF}
\end{equation}
\begin{equation}\label{eq:kappa}
\kappa= \frac{F^2}{RT} \frac{1}{2} D \Bigg(\frac{1}{t_-}+\frac{1}{t_+} \Bigg)c
\end{equation}
\begin{equation}\label{eq:D}
D= \frac{2D_-D_+}{D_+ + D_-}
\end{equation}
\begin{equation}\label{eq:D}
f = \frac{F}{RT}
\end{equation}
\end{subequations}
where the transport parameters, such as $\kappa$ and $D$, are adapted to account for the effects of porosity and tortuosity \cite{verbrugge2005microstructural}. 
The second of the four supercapacitor equations is Ohm's Law, restricted to the electrode domains,
\begin{equation}\label{eq:2}
i_1 = i-i_2 = -\sigma \frac{\partial \Phi_1}{\partial x}.
\end{equation}
The remaining two equations from \cite{verbrugge2005microstructural} are conservation relations, the first for the charge in the electrodes
\begin{equation}\label{eq:3}
aC \frac{\partial (\Phi_1-\Phi_2)}{\partial t}= \frac{\partial i_2}{\partial x},
\end{equation}
and the second being the diffusion equation for the electrolyte concentration
\begin{equation}\label{eq:4}
\epsilon \frac{\partial c}{\partial t}= D \frac{\partial^2 c}{\partial x^2}-\frac{aC}{F}\Bigg( t_- \frac{dq_+}{dq}+t_+\frac{dq_-}{dq} \Bigg) \frac{\partial (\Phi_1-\Phi_2)}{\partial t}.
\end{equation}
The four equations (\ref{eq:1}), (\ref{eq:2}), (\ref{eq:3}) and (\ref{eq:4}) can be written in state-space form as
\begin{gather}\label{eqn:Verburgge_state}
\begin{aligned}
\begin{bmatrix}\epsilon & \frac{aC}{F}(t_-\frac{dq_+}{dq}+t_+\frac{dq_-}{dq}) & -\frac{aC}{F}(t_-\frac{dq_+}{dq}+t_+\frac{dq_-}{dq}) & 0\\
0 & aC & -aC & 0\\
0 & 0 & 0 & 0 \\
0 & 0 & 0 & 0 \end{bmatrix}
\begin{bmatrix} \dot{c} \\ \dot{\Phi}_1 \\ \dot{\Phi}_2 \\ \dot{i}_2\end{bmatrix}
= 
& \begin{bmatrix} D \frac{\partial^2}{\partial x^2} & 0 & 0 & 0\\
0 &0 & 0& \frac{\partial}{\partial x} \\
0 & \sigma \frac{\partial}{\partial x} & 0 &-1 \\
0 & 0 & \kappa \frac{\partial}{\partial x} & 1\end{bmatrix}
\begin{bmatrix} c \\ \Phi_1 \\ \Phi_2 \\i_2\end{bmatrix}
\\
+& \begin{bmatrix} 0 \\0 \\ 0 \\ \kappa \big(\frac{t_+-t_-}{f}\big) \frac{\partial}{\partial x}\end{bmatrix}\text{ln } c +
\begin{bmatrix}0 \\0 \\ 1 \\0 \end{bmatrix}i
\end{aligned}
\end{gather}
in the electrodes and
\begin{gather}\label{eqn:seperator_state}
\begin{aligned}
\begin{bmatrix}\epsilon & 0 \\
0 & 0 \\
\end{bmatrix}
\begin{bmatrix} \dot{c} \\ \dot{\Phi}_2 \\\end{bmatrix}
= 
& \begin{bmatrix} D \frac{\partial^2}{\partial x^2} & 0 \\
0 & \kappa \frac{\partial}{\partial x}\end{bmatrix}
\begin{bmatrix} c \\ \Phi_2 \end{bmatrix}
+& \begin{bmatrix} 0 \\ \kappa \big(\frac{t_+-t_-}{f}\big) \frac{\partial}{\partial x}\end{bmatrix}\text{ln } c +
\begin{bmatrix}0 \\ 1 \end{bmatrix}i
\end{aligned}
\end{gather}
in the separator.

\begin{table}
\centering 
\renewcommand{\arraystretch}{1.3}
\begin{tabular}{|c| c| c|}
\hline 
Parameter &Value & Units \\
\hline
$L_{elect}$ & 50 & $\mu$ m \\
$L_{sep}$ & 25 & $\mu$ m \\
$aC$ & 42 $\times 10^6$ & F/m$^3$ \\
$c_0$ &  930& mol/m$^3$ \\
$\kappa_{\infty}$ &0.067& S/m \\
$t_+$ & 0.5 & \\
$I$ & 100 & A \\
$S$ & 2.747& m$^2$ \\
$i:=I/S$ & 36.403 & A/m$^2$ \\
$\frac{dq_+}{dq}= \frac{dq_-}{dq}$ & -0.5 & \\
$T$ & 298 & K \\
Characteristic Time Constant & 7.4 & s \\
\hline 
\end{tabular}
\caption{Parameters for the supercapacitor model \cite{verbrugge2005microstructural}.}
\label{tab:GlobalParams}
\end{table}

\begin{table}
\centering 
\renewcommand{\arraystretch}{1.3}
\begin{tabular}{|c| c| c|}
\hline
\multicolumn{3}{|c|}{Electrode Region Parameters} \\
\hline 
Parameter &Value & Units \\
\hline
$\epsilon$& 0.67& \\
$\sigma$ & 0.0521 & S/m \\
$\Gamma$ & 2.3& \\
\hline 
\multicolumn{3}{|c|}{Separator Region Parameters} \\
\hline 
Parameter &Value & Units \\
\hline
$\epsilon$& 0.6& \\
$\Gamma$ & 1.29& \\
\hline
\end{tabular}
\caption{Separate parameters for both the separator and electrode regions \cite{verbrugge2005microstructural}.}
\label{tab:ElectSepParams}
\end{table}
The majority of the parameters required by the equations are presented in Tables \ref{tab:GlobalParams} and \ref{tab:ElectSepParams}. The definitions of several others, however, need further discussion. It is assumed in \cite{verbrugge2005microstructural} that the electrolyte concentration $c$ does not vary significantly during the charging profile, so that the relation
\begin{equation}\label{eqn:kappa}
\kappa = \frac{\kappa_{\infty} \epsilon}{\Gamma}
\end{equation}
can be used to calculate the electrolytic conductivity $\kappa$ from the conductivity of the free solution $k_{\infty}$ where the effect of porosity and tortuosity are neglected. Additionally, the total current density flowing through the supercapacitor $i$ is the sum of the current density flowing in the solid phase $i_1$ and the liquid phase $i_2$
\begin{equation}\label{eqn:current}
i_1+i_2 = i.
\end{equation}
In a similar manner, the transference number, a non-dimensional number relating the amount of charge carried by each ion, is defined to sum to 1
\begin{equation}\label{eqn:transference}
t_++t_- = 1.
\end{equation}
The two relations (\ref{eqn:current}) and (\ref{eqn:transference}) can be used to eliminate the variables $i_1$ and $t_-$. The remaining unknown parameter, the ionic diffusion coefficient $D$, can then be defined using (\ref{eq:kappa}), with the concentration $c$ being set to the constant $c_0$ due to the assumption that the concentration does not vary that significantly.

A detailed description of the boundary conditions can be found in \cite{verbrugge2005microstructural}. To summarise, the fluxes at the internal boundary conditions are assumed to be continuous and the total current density flowing through the capacitor $i$ is assumed to be a constant. In the separator region and at the separator/electrode boundary region, all of the current is carried by ions in the electrolyte. Additionally, it is assumed that no ions enter the current collectors, so that the total current at these boundary regions is equal to the solid phase current. 

The four equations (\ref{eq:1}), (\ref{eq:2}), (\ref{eq:3}) and (\ref{eq:4}), can be combined to reduce the number of equations. In this paper, the reduction is achieved by differentiating (\ref{eq:2}) with respect to $x$ 
\begin{equation}
\frac{\partial i_2}{\partial x}= \sigma\frac{\partial^2 \Phi_1}{\partial^2 x},
\end{equation}
and then substituting this expression into (\ref{eq:3})
\begin{equation}
aC\frac{\partial (\Phi_1-\Phi_2)}{\partial t}= \sigma \frac{\partial^2 \Phi_1}{\partial x^2}.
\end{equation}
Summing the two algebraic equations, (\ref{eq:1}) and (\ref{eq:2}), gives
\begin{equation}
0 = \sigma\frac{\partial \Phi_1}{\partial x} + \kappa \frac{\partial \Phi_2}{\partial x}+\kappa \bigg(\frac{t_+-t_-}{f}\bigg)\frac{\partial \text{ ln }c}{\partial x}+i,
\end{equation}
eliminating the algebraic variable $i_2$ from the equation system. This substitution transforms the four equation system of \cite{verbrugge2005microstructural} into the system described by the three equations
\begin{subequations}\label{eqn:DAE2}
\begin{equation}
\epsilon \frac{\partial c}{\partial t}= D \frac{\partial^2 c}{\partial x^2}-\frac{aC}{F}\Bigg( t_- \frac{dq_+}{dq}+t_+\frac{dq_-}{dq} \Bigg) \frac{\partial (\Phi_1-\Phi_2)}{\partial t}
\end{equation}
\begin{equation}
aC\frac{\partial (\Phi_1-\Phi_2)}{\partial t}= \sigma \frac{\partial^2 \Phi_1}{\partial x^2}
\end{equation}
\begin{equation}
0 = \sigma\frac{\partial \Phi_1}{\partial x} + \kappa \frac{\partial \Phi_2}{\partial x}+\kappa \bigg(\frac{t_+-t_-}{f}\bigg)\frac{\partial \text{ ln }c}{\partial x}+i
\end{equation}
\end{subequations}
which for each electrode can be written as
\begin{align}\label{state_space_3}
\begin{split}
\begin{bmatrix}\epsilon & \frac{aC}{F}(t_-\frac{dq_+}{dq}+t_+\frac{dq_-}{dq}) & 0\\
0 & aC & 0 \\
0 & 0 & 0 \end{bmatrix}
\begin{bmatrix} \dot{c} \\ \dot{\Phi}_1-\dot{\Phi}_2 \\ \dot{\Phi}_2\end{bmatrix}
= 
&\begin{bmatrix}D \frac{\partial^2}{\partial x^2} & 0 & 0\\
0 &\sigma \frac{\partial^2}{\partial x^2}&\sigma \frac{\partial^2}{\partial x^2} \\
0& \sigma \frac{\partial}{\partial x} & \kappa \frac{\partial}{\partial x}+ \sigma \frac{\partial}{\partial x} \end{bmatrix}
\begin{bmatrix} c \\ \Phi_1- \Phi_2 \\ \Phi_2\end{bmatrix} 
\\
+& \begin{bmatrix} 0 \\0 \\ \kappa \bigg(\frac{t_+-t_-}{f}\bigg) \frac{\partial}{\partial x}\end{bmatrix}\text{ln }c +
\begin{bmatrix}0 \\0 \\ 1 \end{bmatrix}i.
\end{split}
\end{align}
The state-space representation of the separator domain remains unchanged from (\ref{eqn:seperator_state}). 

A good discussion of several of the assumptions of the model can be found in \cite{johnson1971desalting}. The model uses porous electrode theory, where the electrodes are modelled as one continuum, and is justified as the size of the electrode pores (of the order of nanometres) are much smaller than the electrode thickness (of the order of microns). The electrolyte is considered to be binary, dissociating into two separate ions. If a non-binary electrolyte is used, as in \cite{madabattulamodeling}, then the equation system has to be modified to account for the behaviour of both electrolytic ions. The electrolyte is also assumed to be inert, with the effect of pseudo-capacitance \cite{burke2000ultracapacitors} and side reactions being disregarded. Dilute solution theory, in which the ions are assumed to have zero size, is also used. An investigation into the applicability of this assumption was carried out in \cite{kilic2007steric} where it was found that steric effects between ions due to excessive surface concentrations occurs when the voltage exceeds a certain upper threshold.

One of the most fundamental assumptions of the model equations (\ref{eqn:seperator_state}) and (\ref{state_space_3}) is that the electrolytic concentration does not vary significantly during charging so that the electrolytic conductivity $\kappa$ can be treated as a constant, as described by (\ref{eqn:kappa}).

This assumption can be made more realistic using the substitution
\begin{equation}\label{eqn:beta}
\kappa = \beta c
\end{equation}
that accounts for the linear relationship between electrolyte conductivity and concentration, as mentioned in \cite{verbrugge2005microstructural}. Incorporating (\ref{eqn:beta}) into the model transforms the logarithmic non-linearity  in the algebraic equations of (\ref{eqn:seperator_state}) and (\ref{state_space_3}) into a coupled quadratic non-linearity, as described by
\begin{align}\label{electrode_quadratic}
\begin{split}
&\begin{bmatrix}\epsilon & \frac{aC}{F}(t_-\frac{dq_+}{dq}+t_+\frac{dq_-}{dq}) & 0\\
0 & aC & 0 \\
0 & 0 & 0 \end{bmatrix}
\begin{bmatrix} \dot{c} \\ \dot{\Phi}_1-\dot{\Phi}_2 \\ \dot{\Phi}_2\end{bmatrix}
= \\
&\begin{bmatrix}D \frac{\partial^2}{\partial x^2} & 0 & 0\\
0 &\sigma \frac{\partial^2}{\partial x^2}&\sigma \frac{\partial^2}{\partial x^2} \\
\beta \bigg(\frac{t_+-t_-}{f}\bigg) \frac{\partial}{\partial x}& \sigma \frac{\partial}{\partial x} & \beta c \frac{\partial}{\partial x}+ \sigma \frac{\partial}{\partial x} \end{bmatrix}
\begin{bmatrix} c \\ \Phi_1- \Phi_2 \\ \Phi_2\end{bmatrix} 
+
\begin{bmatrix}0 \\0 \\ 1 \end{bmatrix}i.
\end{split}
\end{align}
in the electrodes and
\begin{gather}\label{eqn:seperator_quadratic}
\begin{aligned}
\begin{bmatrix}\epsilon & 0 \\
0 & 0 \\
\end{bmatrix}
\begin{bmatrix} \dot{c} \\ \dot{\Phi}_2 \\\end{bmatrix}
= 
& \begin{bmatrix} D \frac{\partial^2}{\partial x^2} & 0 \\
\beta \big(\frac{t_+-t_-}{f}\big) \frac{\partial}{\partial x} &\beta c \frac{\partial}{\partial x}\end{bmatrix}
\begin{bmatrix} c \\ \Phi_2 \end{bmatrix}
+
\begin{bmatrix}0 \\ 1 \end{bmatrix}i
\end{aligned}
\end{gather}
in the separator.

\section{Spectral Methods}

In general, non-linear PDE systems such as (\ref{electrode_quadratic}) are too complex to be solved analytically, so numerical methods are used. Numerical methods discretise the spatial domain and then use interpolation to obtain approximate solutions to the PDE, transforming the PDE into a set of ordinary differential equations (ODEs). The manner in which the discretisation is carried out is known to have a significant influence upon the accuracy of the numerical method. The most basic method to increase the accuracy of the numerical solution is to refine the numerical mesh. However, doing so increases the number of ODEs in the model, increasing computational complexity and memory requirements. This results in a trade-off between solution accuracy and computational complexity, with the desired discretisation being such that a low-order model can be established giving a sufficiently accurate solution.

Within numerical methods, the three most common techniques for spatial discretisation are the spectral method (SM), finite difference method (FDM) and finite element method (FEM) \cite{boyd2013chebyshev}. All three of these methods approximate the derivative of an unknown function by differentiating an `approximating function', which is constructed by interpolating the known function values at the domain nodes. The main difference between the FDM, FEM and SM is the domain region used for the interpolation. Both the FDM and FEM can be considered local methods, as they only use information at nodes close to the node of interest where the derivative is to be approximated. This is done either through Taylor expansions for the FDM, or a calculus of variations approach across a sub-domain, known as an element, for the FEM. In contrast, spectral methods represent a global approximation to the derivative, involving a sum of known orthogonal basis functions that traverse the entire domain, with these functions generally chosen to be sinusoids for periodic solutions and polynomials for non-periodic solutions.


Given a suitable spatial grid with a smooth function distributed across it, spectral methods are an accurate numerical method for approximating derivatives. Improving the accuracy in this manner can lead to a significant reduction in computational complexity, as fewer grid points are needed to achieve a set solution accuracy.
The fundamental building block of any spectral method is a set of orthogonal basis functions $\psi$ and in this paper, Chebyshev polynomials of the first kind are used \cite{trefethen2000spectral}. The sum of these basis functions constructs an interpolating polynomial $\bar{p}$ of order $N$ that approximates the smooth function $p$ by
\begin{equation}\label{eq:solution_interp}
p(x) \approx \bar{p}(x) = \sum^{N+1}_{j = 1} \psi_j(x)\bar{p}(\mathbf{x}_j),
\end{equation}
with $p$ being the solution in the spatial domain of the differential equation. The choice of basis function enforces the condition that the approximating polynomial exactly equals the true solution at the collocation points $\mathbf{x}_j$, i.e $\bar{p}(\mathbf{x}_j) = p(\mathbf{x}_j)$.

Approximating the solution to the differential equations through the sum of known polynomial functions in such a manner enables a simple expression for the derivative to be obtained by differentiating the interpolating polynomial 
\begin{equation}\label{eq:diff}
\frac{\partial^l}{\partial x^l}p(x)\approx \frac{\partial^l}{\partial x^l} \bar{p}(x)= \sum^{N+1}_{j=1} \frac{d^l}{dx^l}\psi_j(x)\bar{p}(\mathbf{x}_j).
\end{equation}
The differentiation operation in (\ref{eq:diff}) is linear and can be replaced by a differentiation matrix $\hat{\mathbf{D}}$. The accuracy of the spectral differentiation matrix is exponential \cite{boyd2013chebyshev}, which is superior to the accuracy of the differentiation matrices of both the FDM and FEM. All of the spectral differentiation matrices required for the supercapacitor model proposed in this paper were constructed with the MATLAB differentiation suite \cite{weideman2000matlab}.

\begin{figure}
\centering
\graphicspath{ {Figures/} }
\includegraphics[width=0.75\textwidth]{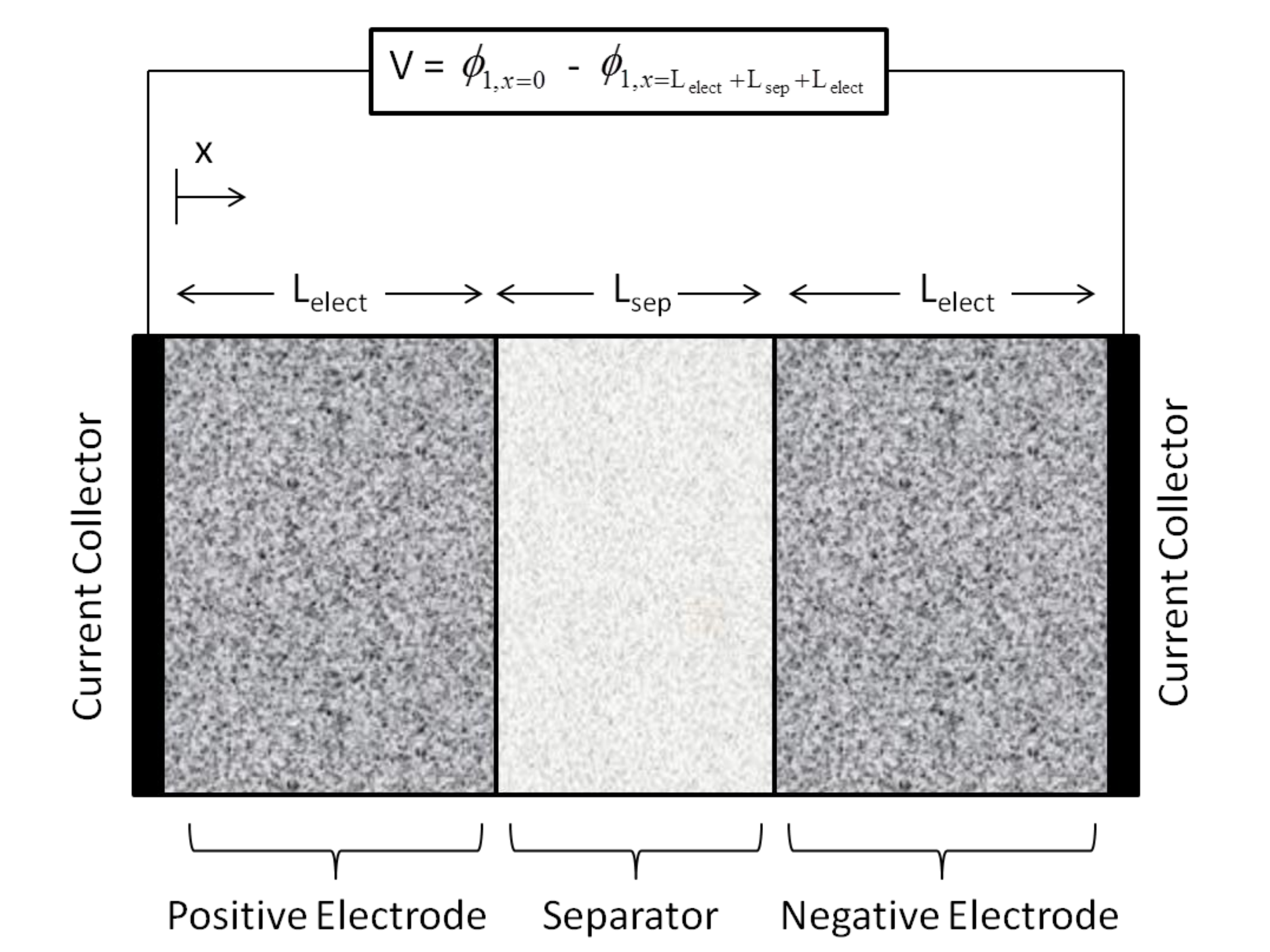}
\caption{Diagram of the supercapacitor showing the electrode and separator regions.}
\label{fig:supercap}
\end{figure}

Spectral methods can give an order of magnitude increase in accuracy for the differentiation of a smooth function. It is known that spectral methods are only applicable for the interpolation of smooth solutions, as the fundamental basis of the method is a global interpolation function, which is itself smooth. For the supercapacitor equations outlined in (\ref{eq:1}), (\ref{eq:2}), (\ref{eq:3}) and (\ref{eq:4}), a discontinuity occurs at the electrode/separator boundary. For this reason, the supercapacitor domain of the presented model is split into three smaller sub-domains, one sub-domain for each of the electrodes and another for the separator, as shown in Figure \ref{fig:supercap}, with the three sub-domains being connected by the boundary conditions using patching \cite{boyd2013chebyshev}. Partitioning the domain in this manner means that the applied spectral method resembles a finite element, albeit one with an interpolating polynomial of very high order. For this reason, the method is commonly known as the spectral element method (SEM).

The accuracy of spectral methods is affected by the Runge phenomenon, where the solution is found to oscillate at the extremal points of the domain as described in \cite{trefethen2000spectral}. This problem can be minimised by discretising the nodes in terms of Chebyshev points $\hat{\mathbf{x}} \in [-1,1]$ 
\begin{equation}\label{eqn:Cheb_Dist}
\hat{\mathbf{x}}_k = \cos \bigg(\frac{(k-1) \pi}{N} \bigg), \quad k = 1,2,3,\dots,N+1,
\end{equation}
with $\hat{\mathbf{x}}$ being the Chebyshev distribution corresponding to the maximal points of the Chebyshev polynomials. The Chebyshev distribution of (\ref{eqn:Cheb_Dist}) is irregular, with a greater density of points being clustered at the domain boundaries, and is defined across the local domain $[-1,1]$. 
For implementation purposes, the transformation from $\hat{\mathbf{x}}$ to $\mathbf{x} \in [0,L_m]$ must be carried out to scale for the true domain \cite{trefethen2000spectral}, and the differentiation matrices must be scaled accordingly.


In this paper, the SEM, FDM and FEM methods were used to discretise the model equations and Neumann boundary conditions of the previous section. The applied FDM uses a second-order Runge-Kutta method as outlined in \cite{verbrugge1994finite}, the SEM implements a similar approach to \cite{bizeray2013advanced}, involving Chebyshev polynomials, while the FEM solutions were obtained using the `Coefficient Form PDE' module from the commercially available software COMSOL.

\section{Differential Algebraic Equations}

The discrete form of the logarithmic non-linear equations given in (\ref{eqn:seperator_state}) and  (\ref{state_space_3}) is respectively given by
\begin{align}\label{discrete_state_space_3}
\begin{split}
\begin{bmatrix}\epsilon & \frac{aC}{F}(t_-\frac{dq_+}{dq}+t_+\frac{dq_-}{dq}) & 0\\
0 & aC & 0 \\
0 & 0 & 0 \end{bmatrix}
\begin{bmatrix} \mathbf{\dot{c}} \\ \mathbf{\dot{\Phi}_1}-\mathbf{\dot{\Phi}_2}\\ \mathbf{\dot{\Phi}_2}\end{bmatrix}
= 
&\begin{bmatrix}D \hat{\mathbf{D}}^2_c & 0 & 0\\
0 &\sigma \hat{\mathbf{D}}^2_{\Phi_1}&\sigma \hat{\mathbf{D}}^2_{\Phi_1} \\
0& \sigma \hat{\mathbf{D}}_{\Phi_1} & \kappa \hat{\mathbf{D}}_{\Phi_2}+\sigma \hat{\mathbf{D}}_{\Phi_1} \end{bmatrix}
\begin{bmatrix} \mathbf{c} \\ \mathbf{\Phi_1}- \mathbf{\Phi_2 }\\ \mathbf{\Phi_2}\end{bmatrix} 
\\
+& \begin{bmatrix} 0 \\0 \\ \kappa \bigg(\frac{t_+-t_-}{f}\bigg) \hat{\mathbf{D}}_{\text{ln }c}\end{bmatrix}\text{ln }\mathbf{c} +
\begin{bmatrix}0 \\ 0 \\ 1 \end{bmatrix}i,
\end{split}
\end{align}
\begin{gather}\label{eqn:seperator_state_discrete}
\begin{aligned}
\begin{bmatrix}\epsilon & 0 \\
0 & 0 \\
\end{bmatrix}
\begin{bmatrix} \dot{\mathbf{c}} \\ \dot{\mathbf{\Phi_2}} \\\end{bmatrix}
= 
& \begin{bmatrix} D \hat{\mathbf{D}}^2_c & 0 \\
0 & \kappa \hat{\mathbf{D}}^2_{\Phi_2}\end{bmatrix}
\begin{bmatrix} \mathbf{c} \\ \mathbf{\Phi_2} \end{bmatrix}
+& \begin{bmatrix} 0 \\ \kappa \big(\frac{t_+-t_-}{f}\big) \hat{\mathbf{D}}_{\text{ln }c}\end{bmatrix}\text{ln } \mathbf{c} +
\begin{bmatrix}0 \\ 1 \end{bmatrix}i
\end{aligned}
\end{gather}
and similarly, the discrete form of the quadratically non-linear model of (\ref{electrode_quadratic}) and (\ref{eqn:seperator_quadratic}) is
\begin{align}\label{electrode_quadratic_discrete}
\begin{split}
&\begin{bmatrix}\epsilon & \frac{aC}{F}(t_-\frac{dq_+}{dq}+t_+\frac{dq_-}{dq}) & 0\\
0 & aC & 0 \\
0 & 0 & 0 \end{bmatrix}
\begin{bmatrix} \mathbf{\dot{c}} \\ \mathbf{\dot{\Phi}_1}-\mathbf{\dot{\Phi}_2}\\ \mathbf{\dot{\Phi}_2} \end{bmatrix}
= \\
&\begin{bmatrix}D \hat{\mathbf{D}}^2_c & 0 & 0\\
0 &\sigma \hat{\mathbf{D}}^2_{\Phi_1}&\sigma \hat{\mathbf{D}}^2_{\Phi_1} \\
\beta  \bigg(\frac{t_+-t_-}{f}\bigg) \hat{\mathbf{D}}_{c}& \sigma \hat{\mathbf{D}}_{\Phi_1} & \beta \mathbf{c} \hat{\mathbf{D}}_{\Phi_2}+ \sigma \hat{\mathbf{D}}_{\Phi_1} \end{bmatrix}
\begin{bmatrix} \mathbf{c} \\ \mathbf{\Phi_1}- \mathbf{\Phi_2 }\\ \mathbf{\Phi_2} \end{bmatrix} 
+
\begin{bmatrix}0 \\0 \\ 1 \end{bmatrix}i,
\end{split}
\end{align}
\begin{gather}\label{eqn:seperator_quadratic_discrete}
\begin{aligned}
\begin{bmatrix}\epsilon & 0 \\
0 & 0 \\
\end{bmatrix}
\begin{bmatrix}  \dot{\mathbf{c}} \\ \dot{\mathbf{\Phi_2}}  \\\end{bmatrix}
= 
& \begin{bmatrix} D \hat{\mathbf{D}}^2_{c} & 0 \\
\beta \big(\frac{t_+-t_-}{f}\big) \hat{\mathbf{D}}_{c} &\beta \mathbf{c} \hat{\mathbf{D}}_{\Phi_2}\end{bmatrix}
\begin{bmatrix} \mathbf{c} \\ \mathbf{\Phi_2}\end{bmatrix}
+
\begin{bmatrix}0 \\ 1 \end{bmatrix}i.
\end{aligned}
\end{gather}
The differentiation matrix $\hat{\mathbf{D}}_c$ includes the boundary conditions on $\mathbf{c}$ and $\hat{\mathbf{D}}_{\Phi_1},\hat{\mathbf{D}}_{\Phi_2},\hat{\mathbf{D}}_{\text{ln }c}$ respectively do the same for $\mathbf{\Phi_1}$, $\mathbf{\Phi_2}$ and ln $\mathbf{c}$.

Equations (\ref{discrete_state_space_3}), (\ref{eqn:seperator_state_discrete}), (\ref{electrode_quadratic_discrete}) and (\ref{eqn:seperator_quadratic_discrete}) are semi-explicit  differential algebraic equations (DAEs), of the form
\begin{subequations}\label{DAE_discrete}
\begin{equation}\label{eqn:Differential}
\mathbf{M} \dot{\mathbf{y}} = \mathbf{f}(\mathbf{y},\mathbf{z}),
\end{equation}
\begin{equation}\label{g}
0 = \mathbf{g}(\mathbf{y},\mathbf{z},\mathbf{u}).
\end{equation}
\end{subequations}
In the electrodes, $\mathbf{y}:=[\mathbf{c},\mathbf{\Phi_1}-\mathbf{\Phi_2}]^T$, $\mathbf{z}:=\mathbf{\Phi_2}$, $\mathbf{u}:=i$,

\begin{equation}
\mathbf{M}:=\begin{bmatrix}\epsilon & \frac{aC}{F}(t_-\frac{dq_+}{dq}+t_+\frac{dq_-}{dq}) \\
0 & aC  \end{bmatrix},
\end{equation}
\begin{equation}
 \mathbf{f}(\mathbf{y},\mathbf{z}):= \begin{bmatrix}D \hat{\mathbf{D}}^2_c & 0 & 0\\
0 &\sigma \hat{\mathbf{D}}^2_{\Phi_1}&\sigma \hat{\mathbf{D}}^2_{\Phi_1} \end{bmatrix}
\begin{bmatrix} \mathbf{c} \\ \mathbf{\Phi_1}- \mathbf{\Phi_2 }\\ \mathbf{\Phi_2} \end{bmatrix} 
\end{equation}
and the algebraic equation $\mathbf{g}(\mathbf{y},\mathbf{z},\mathbf{u})$ is respectively defined for the logarithmic and quadratic models as
\begin{equation}
\mathbf{g}(\mathbf{y},\mathbf{z},\mathbf{u}):= 
\begin{bmatrix} 0& \sigma \hat{\mathbf{D}}_{\Phi_1} & \kappa \hat{\mathbf{D}}_{\Phi_2}+\sigma \hat{\mathbf{D}}_{\Phi_1} \end{bmatrix}
\begin{bmatrix} \mathbf{c} \\ \mathbf{\Phi_1}- \mathbf{\Phi_2 }\\ \mathbf{\Phi_2}\end{bmatrix} 
\\
+ \kappa \bigg(\frac{t_+-t_-}{f}\bigg) \hat{\mathbf{D}}_{\text{ln }c}\text{ln }\mathbf{c} +i,
\end{equation}
\begin{equation}
\mathbf{g}(\mathbf{y},\mathbf{z},\mathbf{u}):= \begin{bmatrix}\beta  \bigg(\frac{t_+-t_-}{f}\bigg) \hat{\mathbf{D}}_{c}& \sigma \hat{\mathbf{D}}_{\Phi_1} & \beta \mathbf{c} \hat{\mathbf{D}}_{\Phi_2}+ \sigma \hat{\mathbf{D}}_{\Phi_1} \end{bmatrix}
\begin{bmatrix} \mathbf{c} \\ \mathbf{\Phi_1}- \mathbf{\Phi_2 }\\ \mathbf{\Phi_2} \end{bmatrix} 
+i.
\end{equation}
Equivalently, in the separator, $\mathbf{y}:=\mathbf{c}$, $\mathbf{z}:=\mathbf{\Phi_2}$, $\mathbf{u}:=i$,
\begin{equation}
\mathbf{M}:= \epsilon,
\end{equation}
\begin{equation}
 \mathbf{f}(\mathbf{y},\mathbf{z}):= \begin{bmatrix} D \hat{\mathbf{D}}^2_c & 0 \end{bmatrix}
\begin{bmatrix} \mathbf{c} \\ \mathbf{\Phi_2} \end{bmatrix}
\end{equation}
 and $\mathbf{g}(\mathbf{y},\mathbf{z},\mathbf{u})$ is respectively defined according to
\begin{equation}
\mathbf{g}(\mathbf{y},\mathbf{z},\mathbf{u}):= 
\begin{bmatrix} 0& \kappa \hat{\mathbf{D}}_{\Phi_2} \end{bmatrix}
\begin{bmatrix} \mathbf{c} \\  \mathbf{\Phi_2}\end{bmatrix} 
\\
+ \kappa \bigg(\frac{t_+-t_-}{f}\bigg) \hat{\mathbf{D}}_{\text{ln }c}\text{ln }\mathbf{c} +i,
\end{equation}
\begin{equation}
\mathbf{g}(\mathbf{y},\mathbf{z},\mathbf{u}):= \begin{bmatrix}\beta  \bigg(\frac{t_+-t_-}{f}\bigg) \hat{\mathbf{D}}_{c}&  \beta \mathbf{c} \hat{\mathbf{D}}_{\Phi_2} \end{bmatrix}
\begin{bmatrix} \mathbf{c} \\ \mathbf{\Phi_2} \end{bmatrix} 
+i,
\end{equation}
for the logarithmic and quadratic models. 

A key parameter for solving any DAE system is its index, defined as the number of derivatives needed to transform the DAE into an ODE. DAEs of index 1 are of particular interest, as they are significantly simpler to solve. To solve an index 1 DAE system, the Jacobian of the algebraic equation (\ref{g}) is taken according to
\begin{equation}
0 = \frac{\partial \mathbf{g}}{\partial \mathbf{y}} \mathbf{y}+\frac{\partial \mathbf{g}}{\partial \mathbf{z}} \mathbf{z} + \frac{\partial \mathbf{g}}{\partial \mathbf{u}} \mathbf{u},
\end{equation}
leading to an expression for the algebraic variable $\mathbf{z}$ 
\begin{equation}
\mathbf{z} = -\Bigg[ \frac{\partial \mathbf{g}}{\partial \mathbf{z}}\Bigg]^{-1}\Bigg( \frac{\partial \mathbf{g}}{\partial \mathbf{y}} \mathbf{y}+ \frac{\partial \mathbf{g}}{\partial \mathbf{u}} \mathbf{u}\Bigg)
\end{equation}
that can be obtained provided that $\partial \mathbf{g}/ \partial \mathbf{z}$ is non-singular. Substituting this expression for $\mathbf{z}$ into (\ref{eqn:Differential}) transforms $\mathbf{f}$ into a function of $\mathbf{y}$ and $\mathbf{u}$ only
\begin{equation}\label{eqn:DAE}
\mathbf{M} \dot{\mathbf{y}} = \mathbf{f}(\mathbf{y},\mathbf{u}).
\end{equation}

For the logarithmically non-linear model of (\ref{discrete_state_space_3}) and (\ref{eqn:seperator_state_discrete}), the derivative of $\mathbf{g}$ with respect to the algebraic variable $\mathbf{\Phi_2}$ in the electrodes is given by
\begin{equation} \label{invertible_elect}
\frac{\partial \mathbf{g}}{\partial \mathbf{z}} = \kappa \hat{\mathbf{D}}_{\Phi_2}+ \sigma \hat{\mathbf{D}}_{\Phi_1}
\end{equation}
and in the separator by
\begin{equation} \label{invertible_sep}
\frac{\partial \mathbf{g}}{\partial \mathbf{z}} = \kappa \hat{\mathbf{D}}_{\Phi_2}.
\end{equation}
The derivative  $\partial \mathbf{g}/ \partial \mathbf{z}$ for the quadratically non-linear model of (\ref{electrode_quadratic_discrete}) and (\ref{eqn:seperator_quadratic_discrete}) is given by
\begin{equation} \label{invertible_elect_quad}
\frac{\partial \mathbf{g}}{\partial \mathbf{z}} = \beta \mathbf{c} \hat{\mathbf{D}}_{\Phi_2}+ \sigma \hat{\mathbf{D}}_{\Phi_1}
\end{equation}
in the electrodes and by
\begin{equation} \label{invertible_sep_quad}
\frac{\partial \mathbf{g}}{\partial \mathbf{z}} = \beta \mathbf{c} \hat{\mathbf{D}}_{\Phi_2}.
\end{equation}
 in the separator.
The matrices (\ref{invertible_elect}), (\ref{invertible_sep}), (\ref{invertible_elect_quad}) and (\ref{invertible_sep_quad}) are all invertible, and as such (\ref{discrete_state_space_3}),  (\ref{eqn:seperator_state_discrete}), (\ref{eqn:seperator_quadratic_discrete}) and (\ref{electrode_quadratic_discrete})  are DAEs of index 1. This means that the models' initial value problem can be integrated using a solver such as MATLAB's ode15s routine, which uses a Newton iteration method to solve the algebraic equations and an implicit numerical differentiation formula (NDF) to carry out the integration \cite{shampine1997matlab, shampine1999solving}.

\section{Results}

Experimental data of typical charging profiles of a supercapacitor from SAFT America \cite{chu2002comparison} is presented in \cite{verbrugge2005microstructural}.

Most of the parameters of the supercapacitor are given in Tables \ref{tab:GlobalParams} and \ref{tab:ElectSepParams}, with the rest being calculated using equations (\ref{eqn:kappa}), (\ref{eqn:current}) and (\ref{eqn:transference}).
In \cite{verbrugge2005microstructural}, the supercapacitor was first charged from an initial voltage of 1.63 V at a constant current of 100 A for 23.2 s whereupon the voltage was then held for 6 s at a constant value of 1.41 V, a constant-current, constant-voltage (CC-CV) charging profile. In this paper, this CC-CV profile is labelled the standard charging profile. Constant current charging profiles with currents up to 1000 A were also simulated to show that the models could accommodate high currents.

\begin{figure}
\centering
\graphicspath{ {Figures/} }
\begin{subfigure}[b]{0.5\textwidth}
\includegraphics[width=\textwidth]{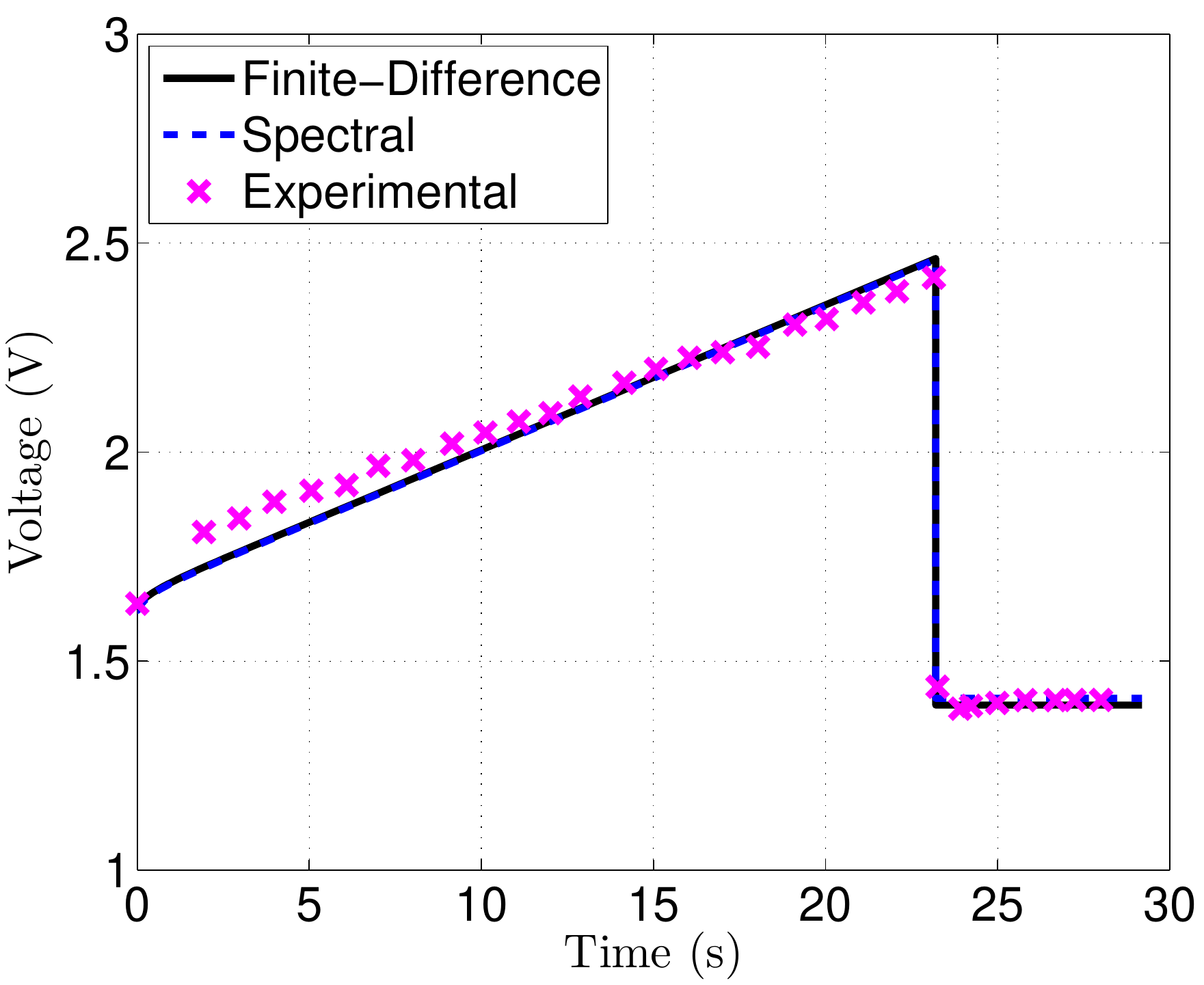}
\caption{Voltage.}
\label{fig:solvers_C}
\end{subfigure}
\\
\hspace{-18pt}
\begin{subfigure}[b]{0.53\textwidth}
\includegraphics[width=\textwidth]{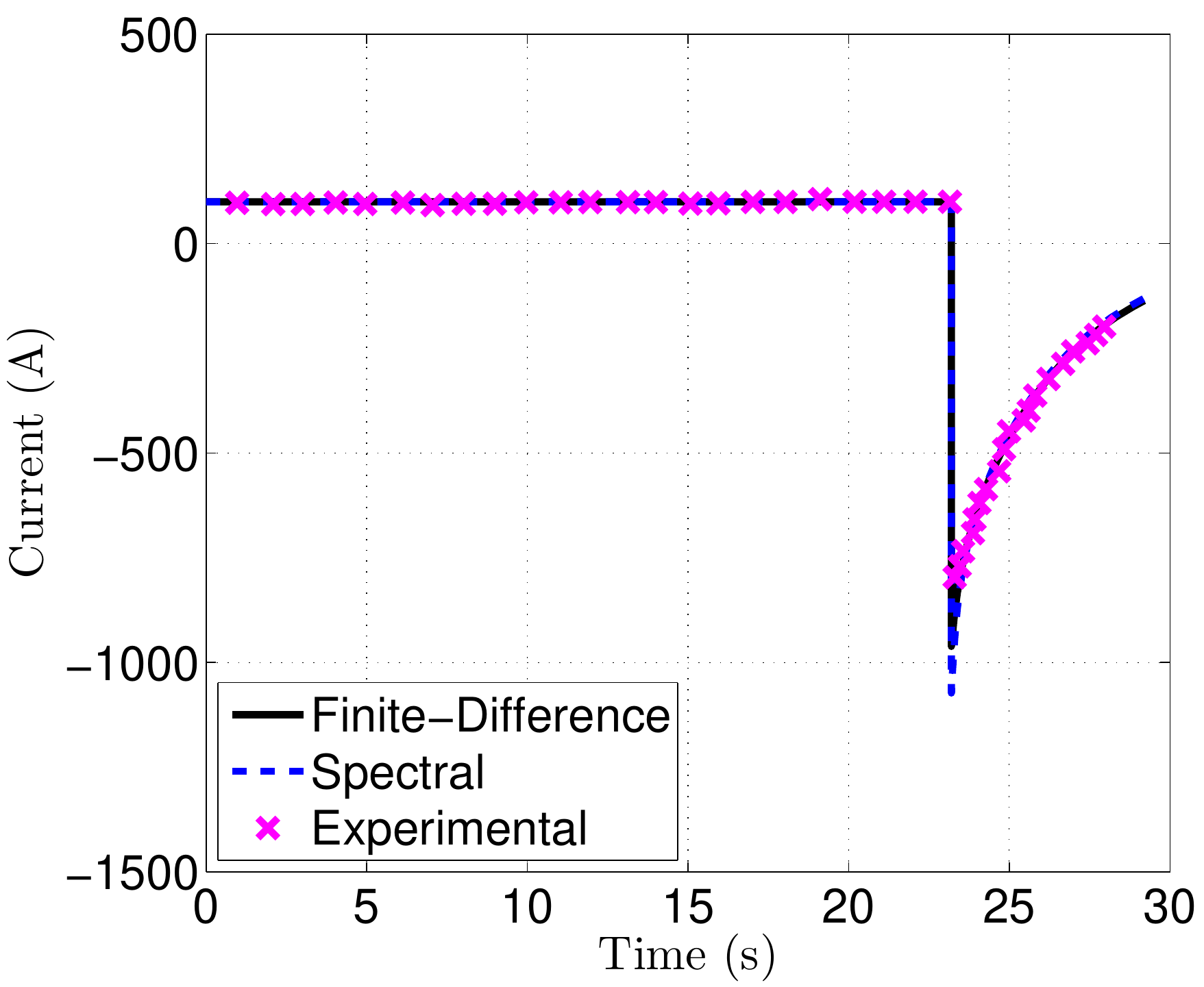}
\caption{Current.}
\label{fig:solvers_V}
\end{subfigure}
\caption{The voltage/current responses of  (\ref{discrete_state_space_3}) and (\ref{eqn:seperator_state_discrete}) for the standard charging profile discretised using the FDM and SEM with the experimental data of \cite{verbrugge2005microstructural}.}
\label{fig:Solvers_output}
\end{figure}

Figure \ref{fig:Solvers_output} compares the model outputs from (\ref{discrete_state_space_3}) and (\ref{eqn:seperator_state_discrete}) solved using the SEM and FDM with the experimental data presented in \cite{verbrugge2005microstructural}. 

Both models can be seen to match well with the experimental data, validating the model assumptions for this typical charging profile. 
Furthermore, the size of the FDM and SEM models are small, using only 5 elements in each domain. %
This contrasts with the generalised finite element method of \cite{allu2014generalized}, which also compares its results against the data of \cite{verbrugge2005microstructural}, that used 1200 and 2500 finite elements. 
This shows that accurate results can be obtained using low order models. 

\begin{figure}
\centering
\graphicspath{ {Figures/} }
\begin{subfigure}{0.49\textwidth}
\includegraphics[width=1\textwidth]{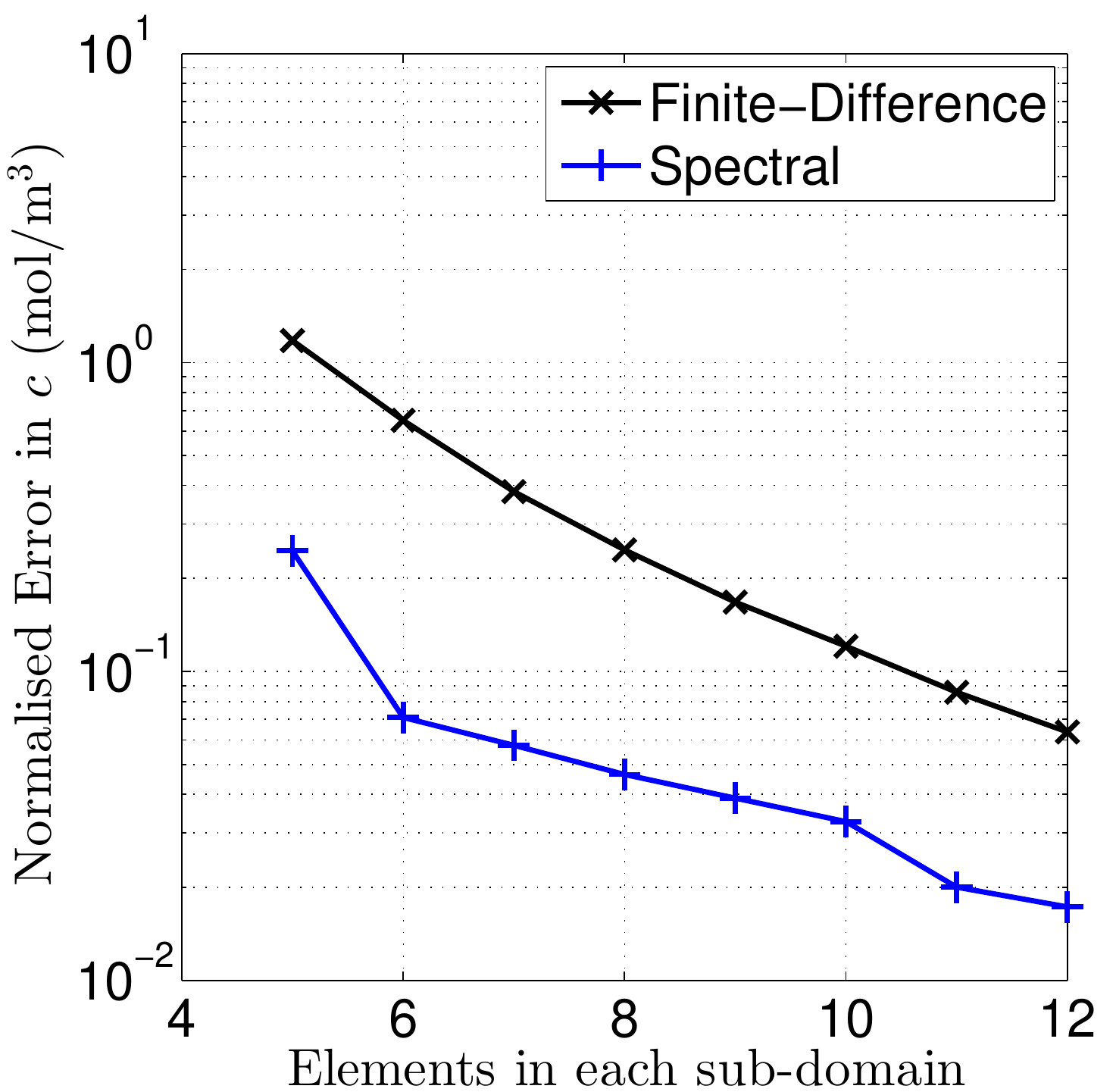}
\caption{Concentration $c$.}
\label{fig:con_C}
\end{subfigure}
\\
\begin{subfigure}[b]{0.49\textwidth}
\includegraphics[width=1\textwidth]{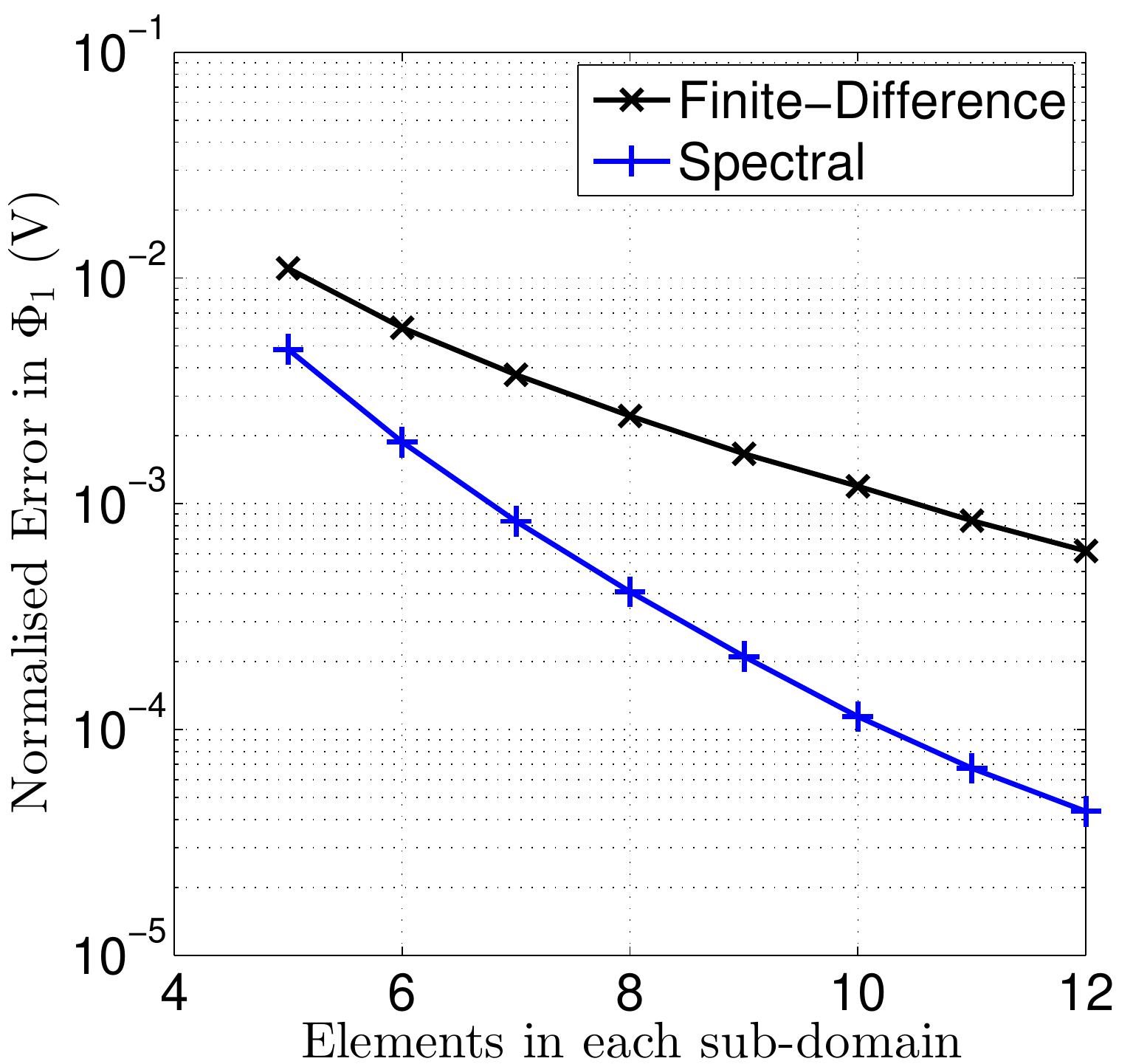}
\caption{Electrode Potential $\Phi_1$.}
\label{fig:con_phi1}
\end{subfigure}
\begin{subfigure}[b]{0.49\textwidth}
\includegraphics[width=1\textwidth]{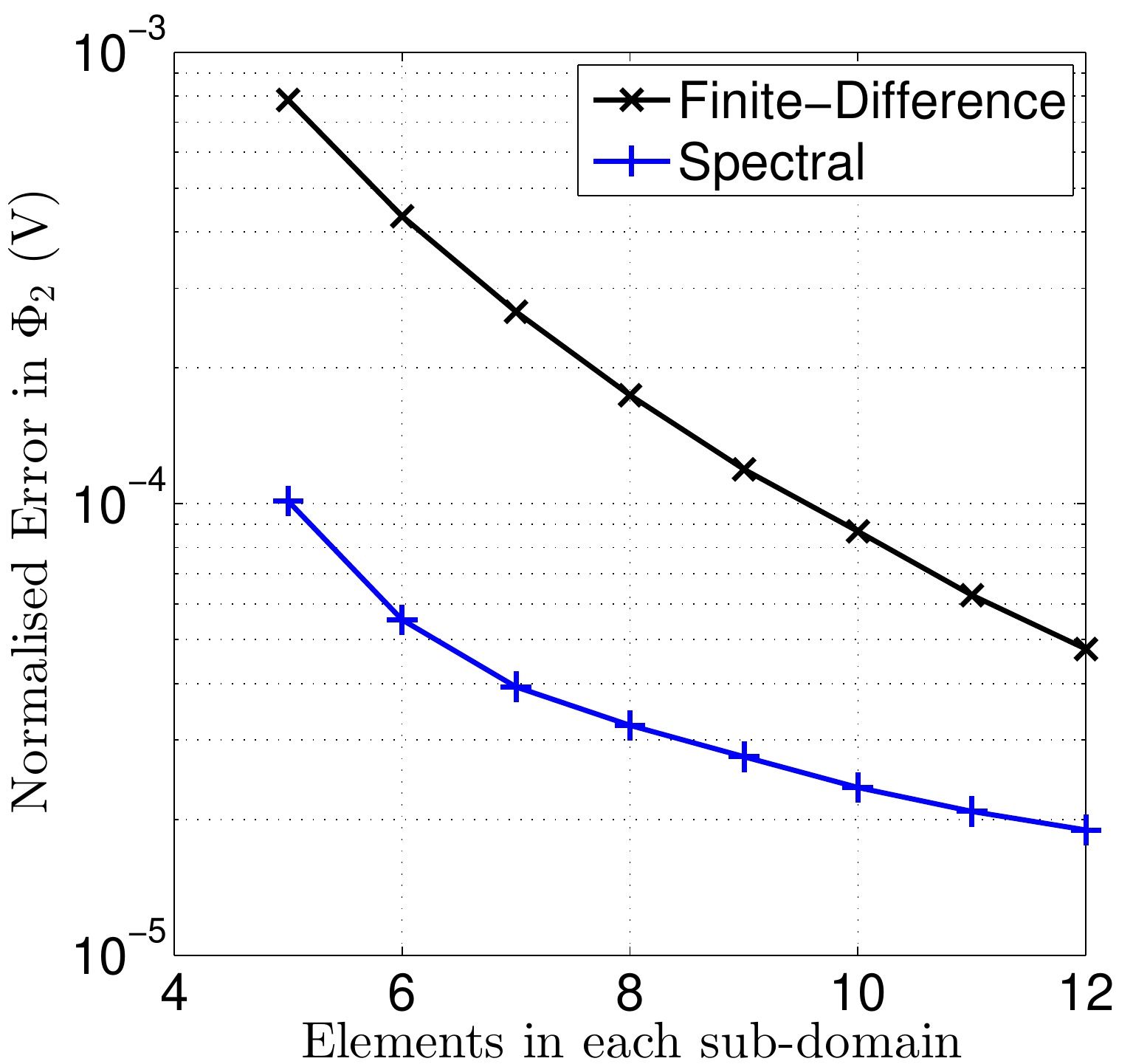}
\caption{Electrolyte Potential $\Phi_2$.}
\label{fig:con_phi2}
\end{subfigure}
\caption{The normalised errors of (\ref{discrete_state_space_3}) and (\ref{eqn:seperator_state_discrete}) discretised using the SEM and FDM, with the reference solution being obtained from COMSOL.}
\label{fig:convergence}
\end{figure}

The accuracy of the model states, instead of the model outputs, is investigated in Figure \ref{fig:convergence}. In this figure, the error convergence rates with respect to the number of elements in each domain for the logarithmic model (\ref{discrete_state_space_3}) and (\ref{eqn:seperator_state_discrete}) discretised using the FDM and SEM are shown.

\begin{figure}
\centering
\graphicspath{ {Figures/} }
\includegraphics[width=0.6\textwidth]{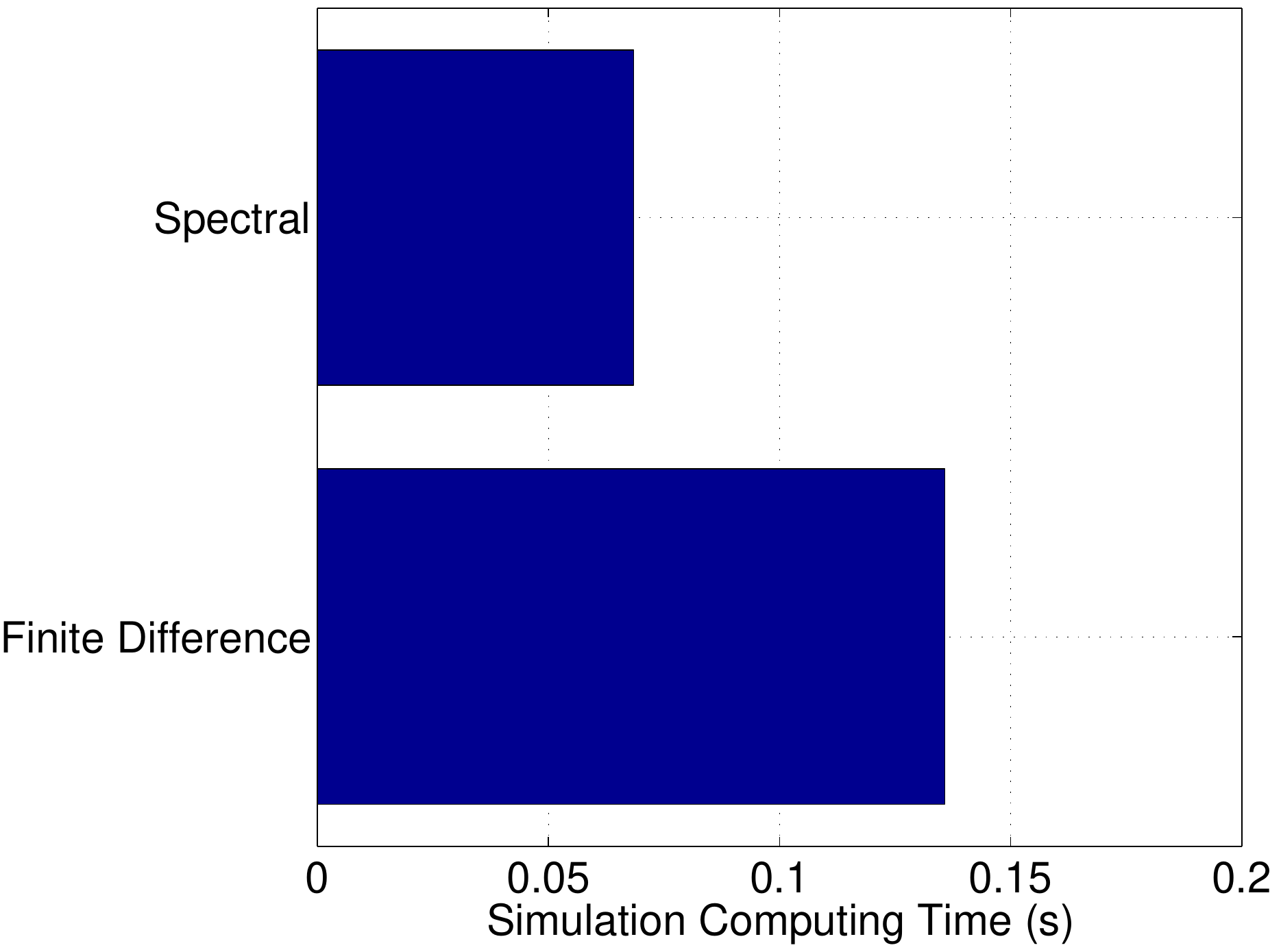}
\caption{Simulation computing times for the standard charging profile of (\ref{discrete_state_space_3}) and (\ref{eqn:seperator_state_discrete}) discretised using the FDM and SEM. Computing times were recorded using the MATLAB `tictoc' command.}
\label{fig:timing}
\end{figure}

For this comparison, the FEM solution with a high number of nodes (41 in each electrode and 18 in the separator) obtained from COMSOL is taken to be the reference numerical solution.
The error of the figure is defined as the 2-norm of the absolute error between the solver and the reference solutions, normalised with respect to the number of time steps and spatial elements. Figure \ref{fig:convergence} shows that the error of the SEM converges much faster than the FDM, indicating that a given level of accuracy can be obtained with fewer nodes using the SEM. 
In Figure \ref{fig:timing}, simulation computing times for the standard charging profile of (\ref{discrete_state_space_3}) and (\ref{eqn:seperator_state_discrete}) discretised using the FDM and SEM are recorded using the MATLAB `tictoc' command. The solution accuracies for both simulations are intended to be kept approximately the same and this is achieved by respectively using 6 and 12 elements in each domain for the SEM and FDM discretised models. 
It is shown that, for a similar level of solution accuracy, the simulation computing time of the SEM model is approximately 52\% that of the FDM model.
This implies that the SEM discretised model would be superior when used for accurate state estimation with an observer and as the basis for an online controller.

\begin{figure}
\centering
\graphicspath{ {Figures/} }
\begin{subfigure}[b]{0.5\textwidth}
\includegraphics[width=\textwidth]{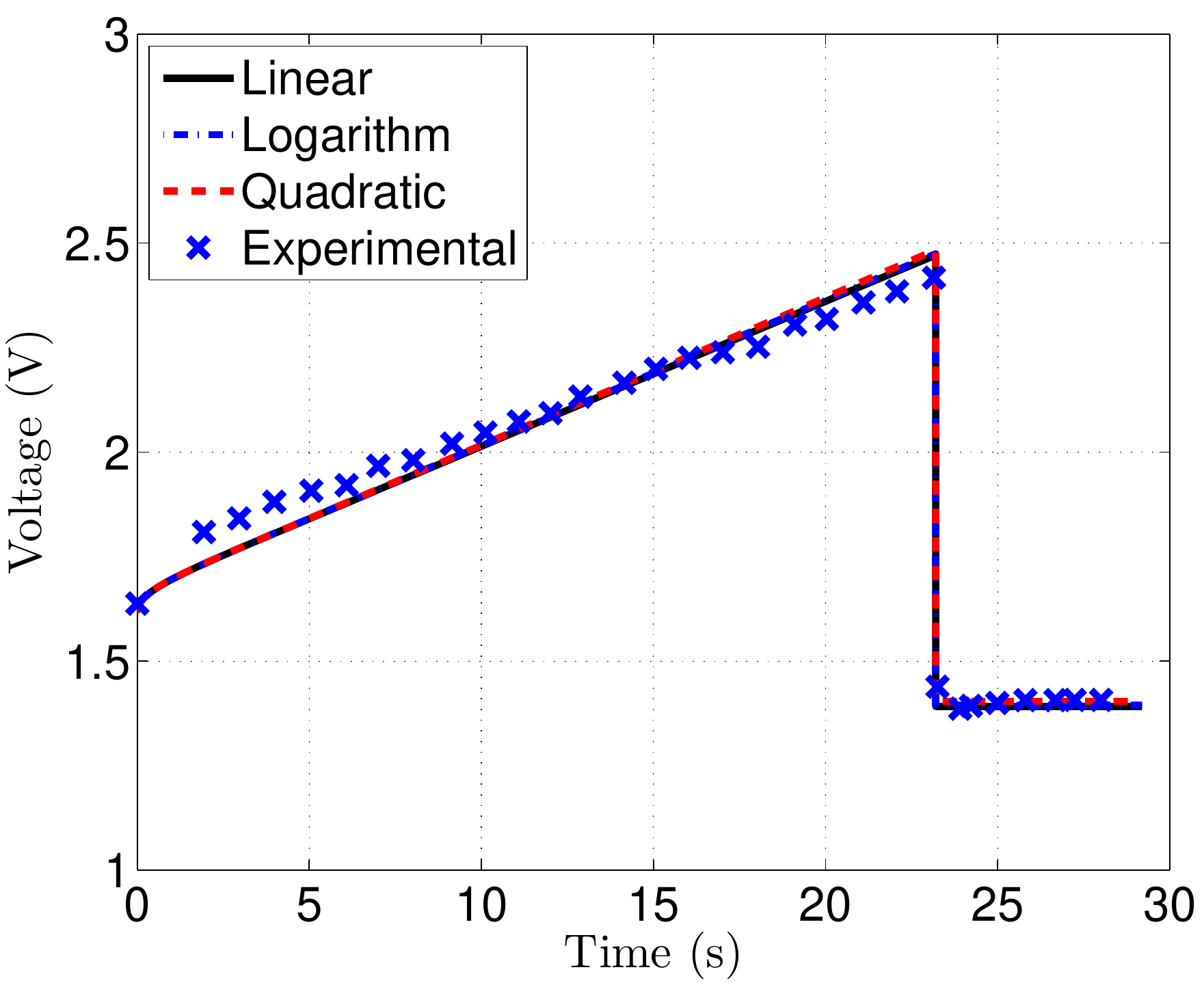}
\caption{Voltage.}
\label{fig:models_C}
\end{subfigure}
\\
\hspace{-18pt}
\begin{subfigure}[b]{0.53\textwidth}
\includegraphics[width=\textwidth]{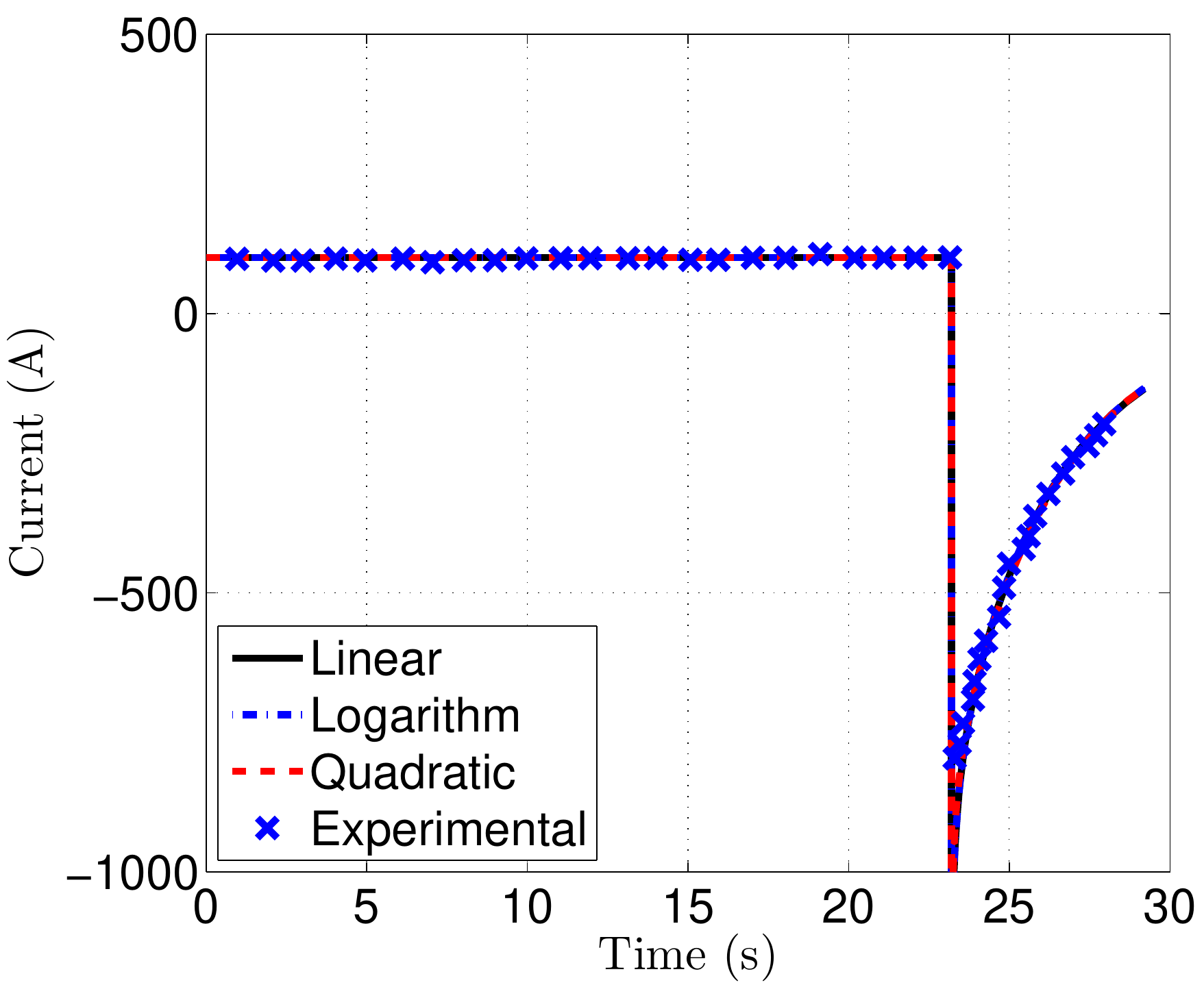}
\caption{Current.}
\label{fig:models_V}
\end{subfigure}
\caption{The voltage/current response of the non-linear models and experimental data of \cite{verbrugge2005microstructural} for the standard CC-CV charging profile.}
\label{fig:output_Verb}
\end{figure}

\begin{figure}
\centering
\graphicspath{ {Figures/} }
\hspace{0pt}
\begin{subfigure}[b]{0.45\textwidth}
\includegraphics[width=\textwidth]{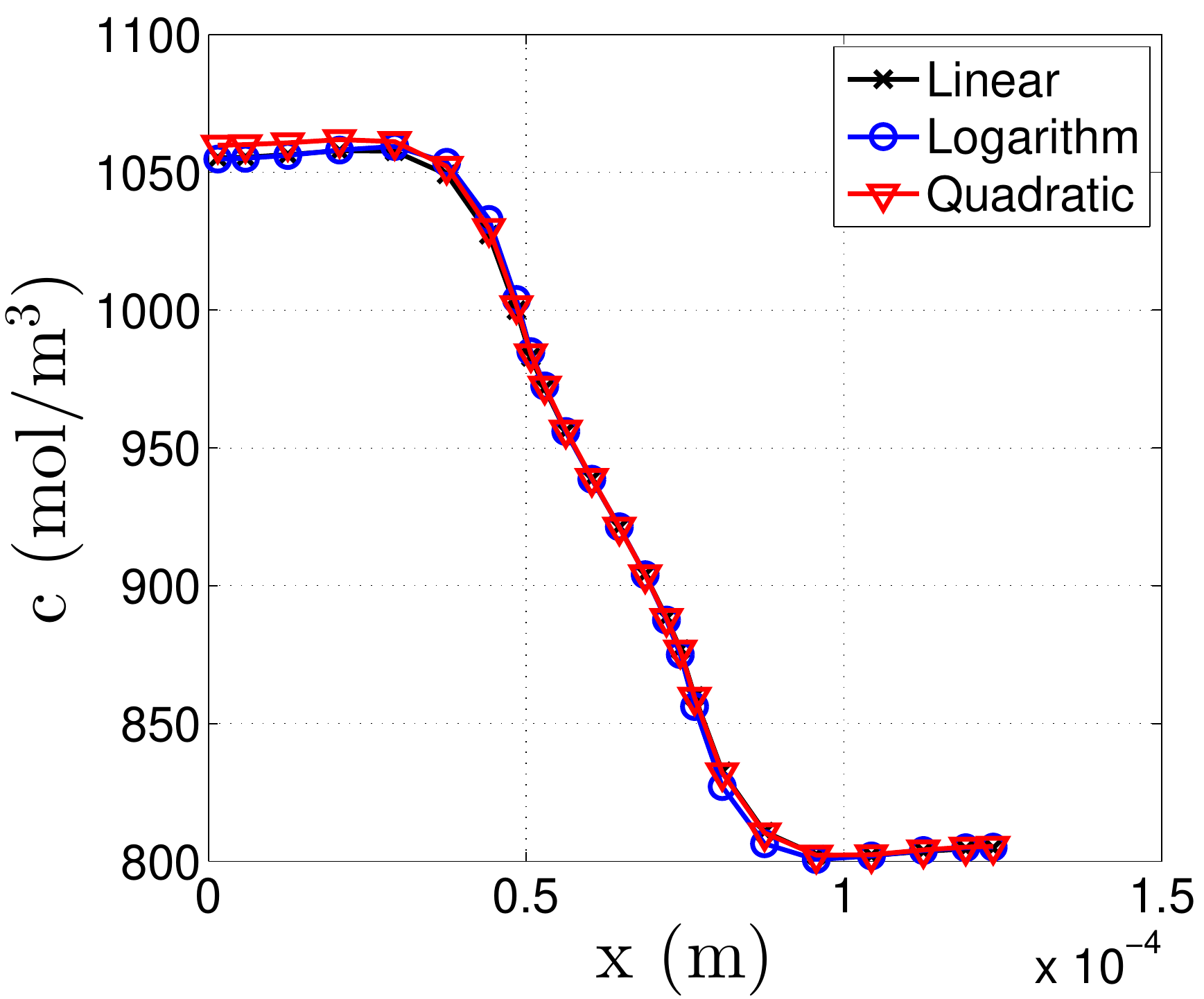}
\caption{Electrolyte concentration.}
\label{fig:states_Verb_c}
\end{subfigure}
\quad
\hspace{-4pt}
\begin{subfigure}[b]{0.472\textwidth}
\includegraphics[width=1\textwidth]{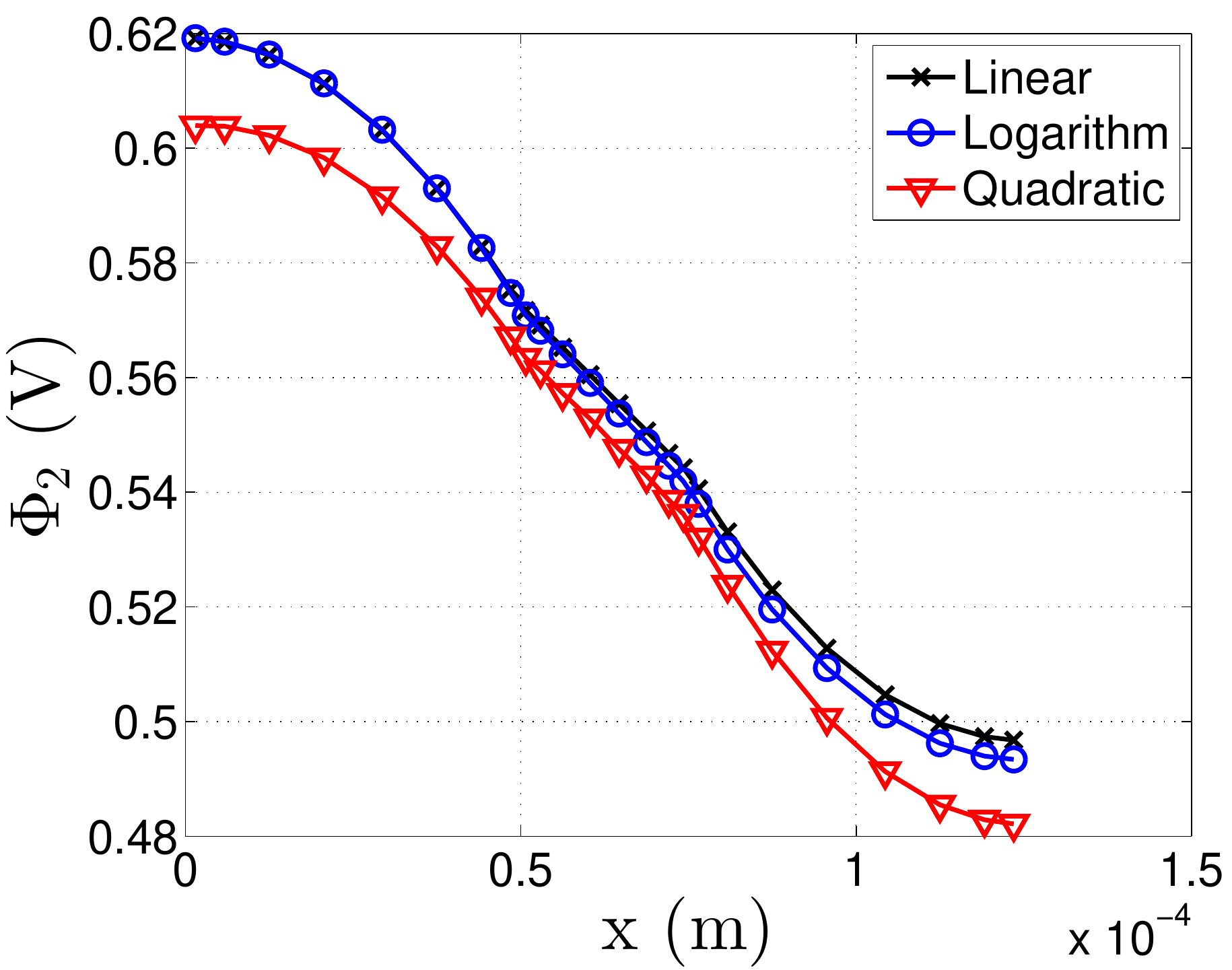}
\caption{Electrolyte potential.}
\label{fig:states_Verb_phi2}
\end{subfigure}
\\
\hspace{2pt}
\hspace{-15pt}
\begin{subfigure}[b]{0.5\textwidth}
\includegraphics[width=\textwidth]{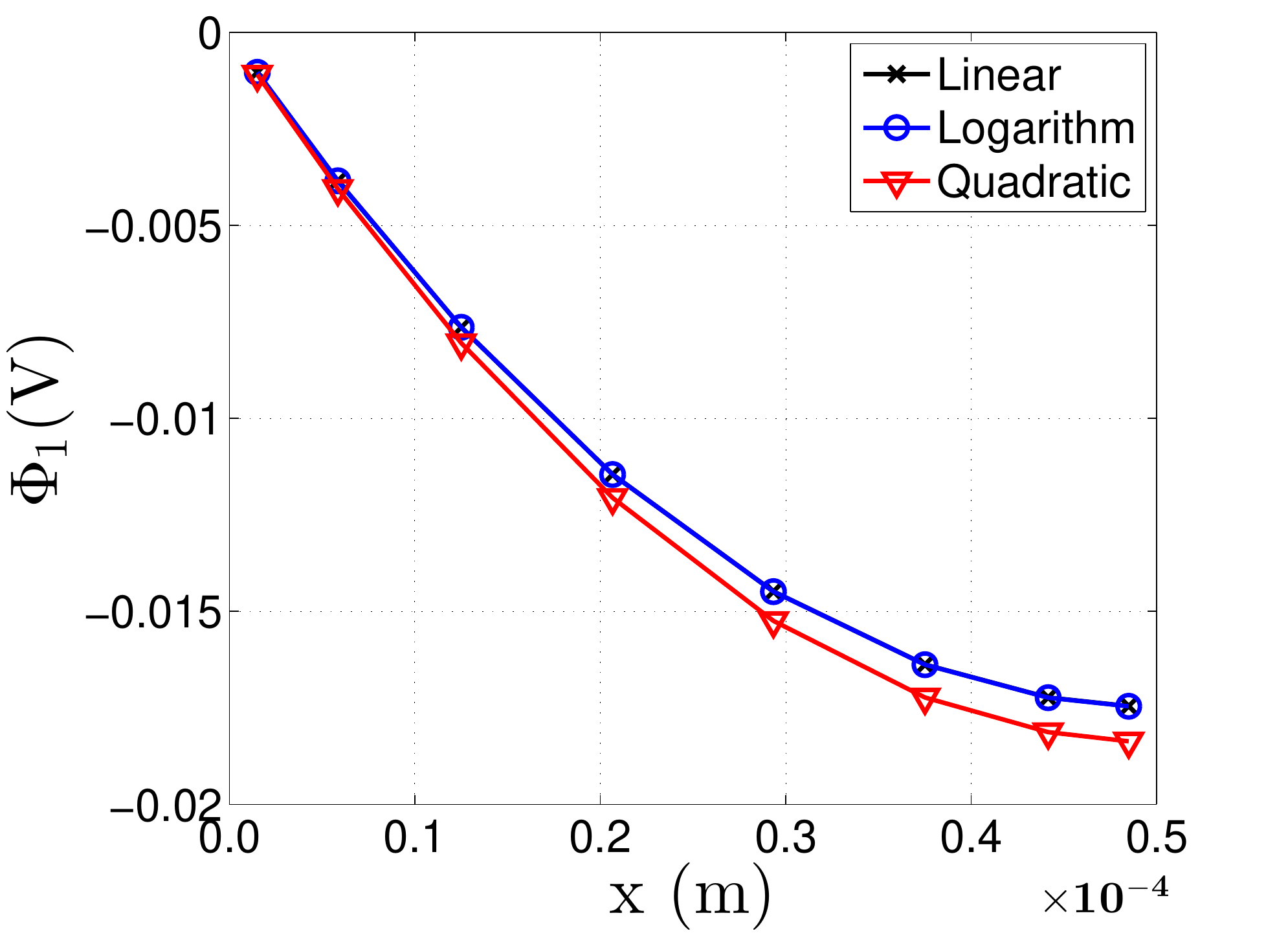}
\caption{Potential in the left electrode.}
\label{fig:states_Verb_phiL}
\end{subfigure}
\hspace{-6pt}
\begin{subfigure}[b]{0.45\textwidth}
\includegraphics[width=\textwidth]{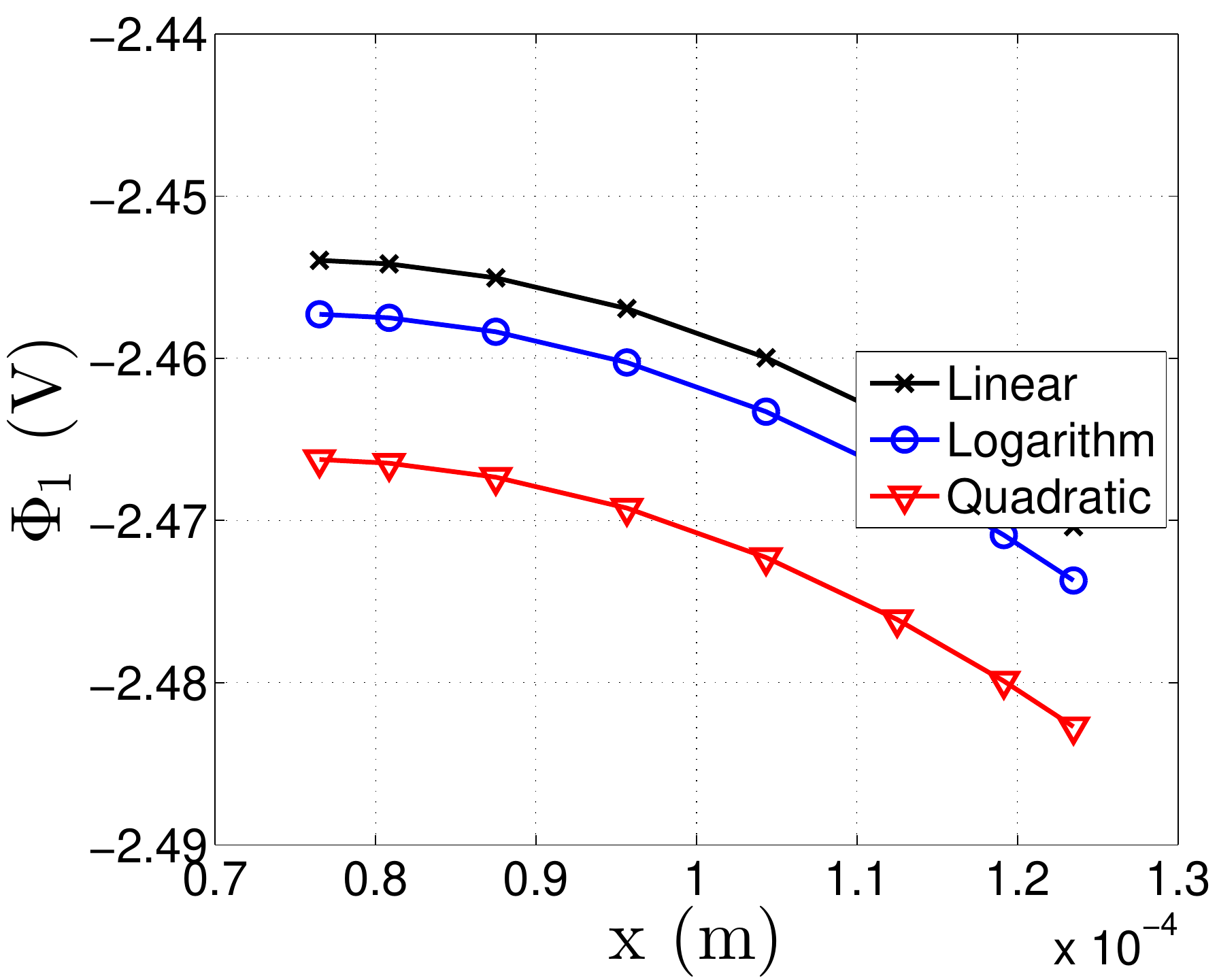}
\caption{Potential in the right electrode.}
\label{fig:states_Verb_phiR}
\end{subfigure}
\caption{The state distribution of the non-linear models after 23.2 s  of the standard charging profile.}
\label{fig:states_Verb}
\end{figure}

\begin{figure}
\centering
\graphicspath{ {Figures/} }
\hspace{-1pt}
\begin{subfigure}[b]{0.45\textwidth}
\includegraphics[width=\textwidth]{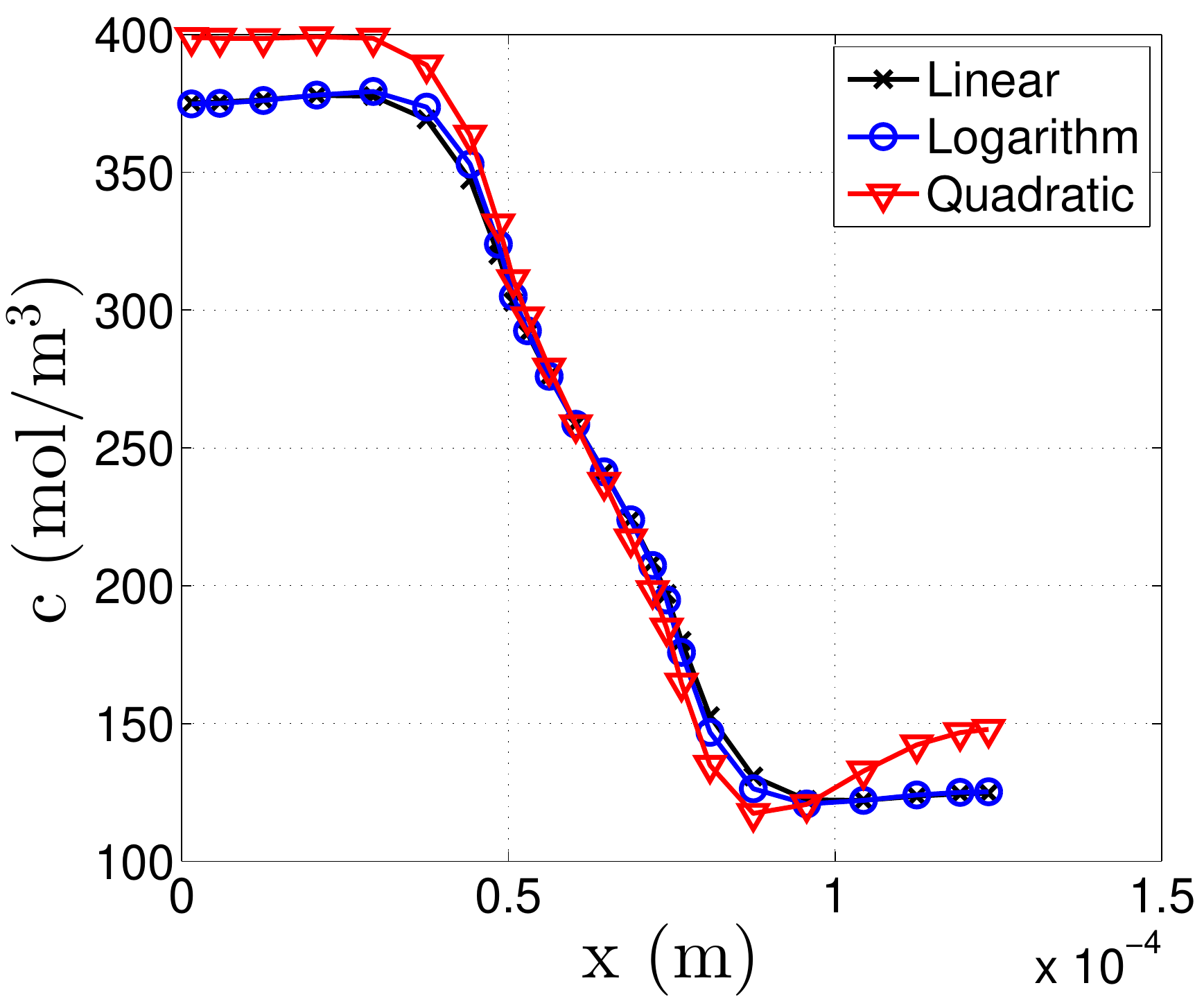}
\caption{Electrolyte concentration.}
\label{fig:states_Dilute_c}
\end{subfigure}
\quad
\hspace{2pt}
\begin{subfigure}[b]{0.475\textwidth}
\includegraphics[width=\textwidth]{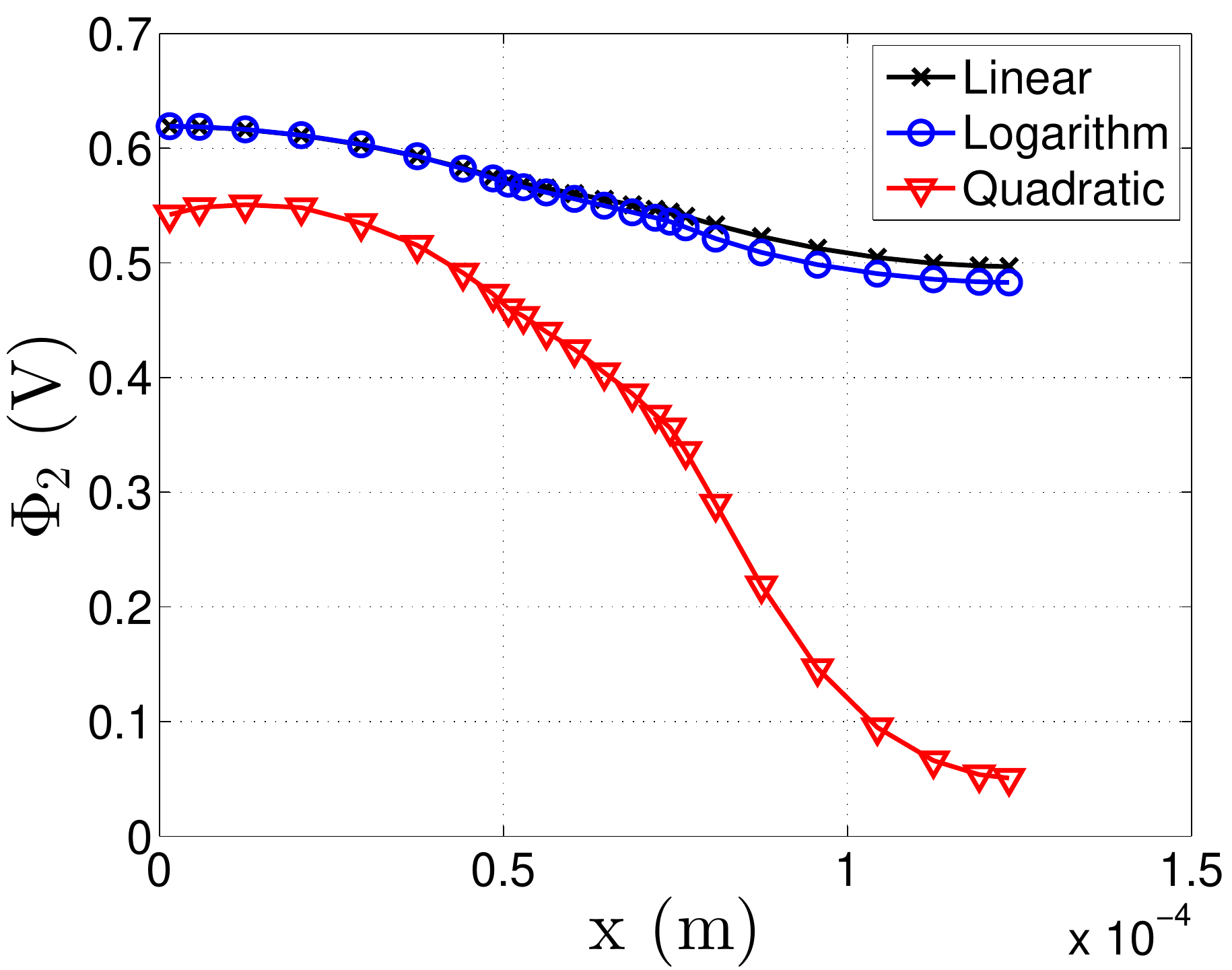}
\caption{Electrolyte potential.}
\label{fig:states_Dilute_phi2}
\end{subfigure}
\\
\hspace{2pt}
\hspace{-17pt}
\begin{subfigure}[b]{0.51\textwidth}
\includegraphics[width=\textwidth]{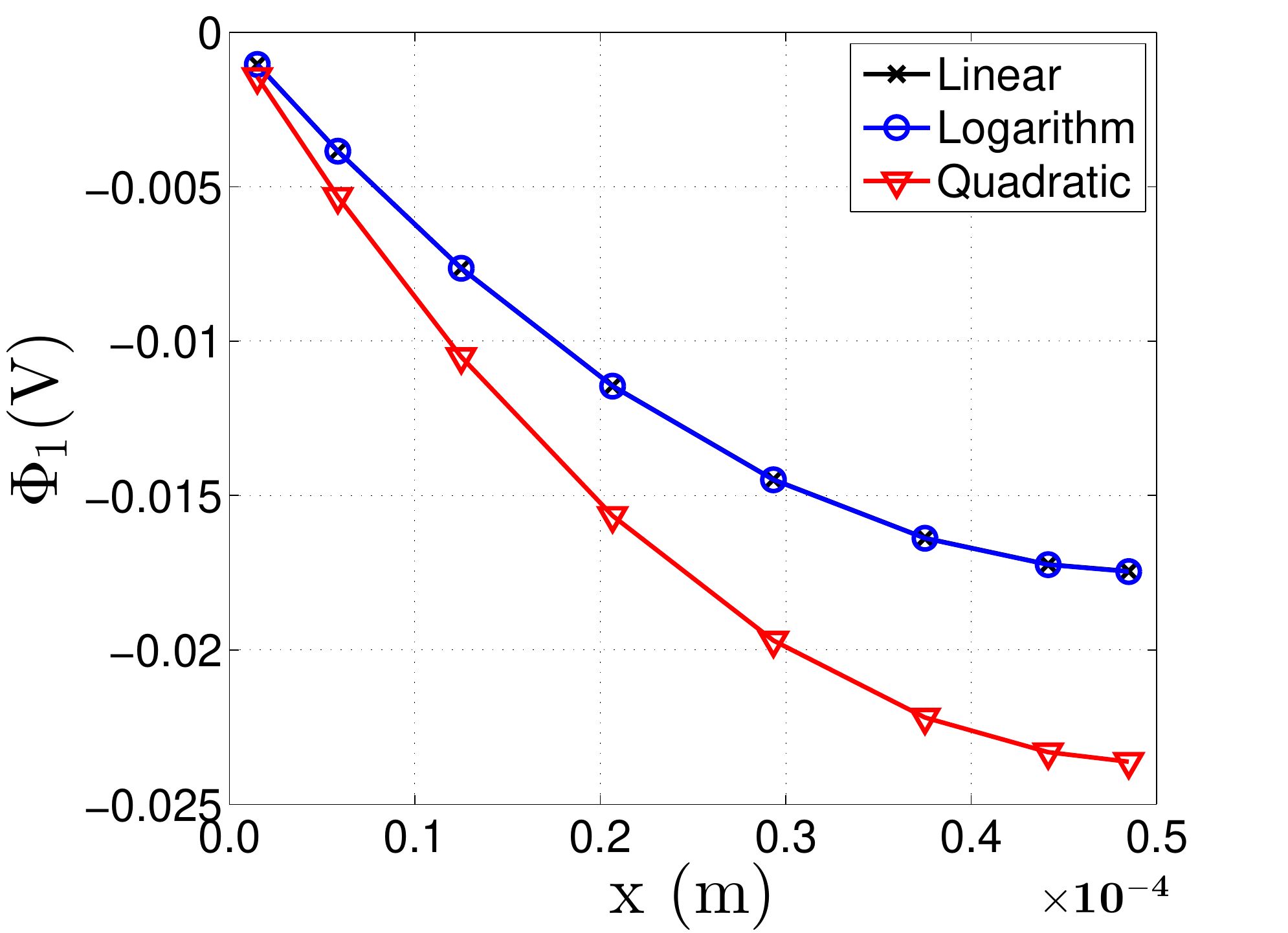}
\caption{Potential in the left electrode.}
\label{fig:states_Dilute_phi1R}
\end{subfigure}
\hspace{-6pt}
\begin{subfigure}[b]{0.47\textwidth}
\includegraphics[width=\textwidth]{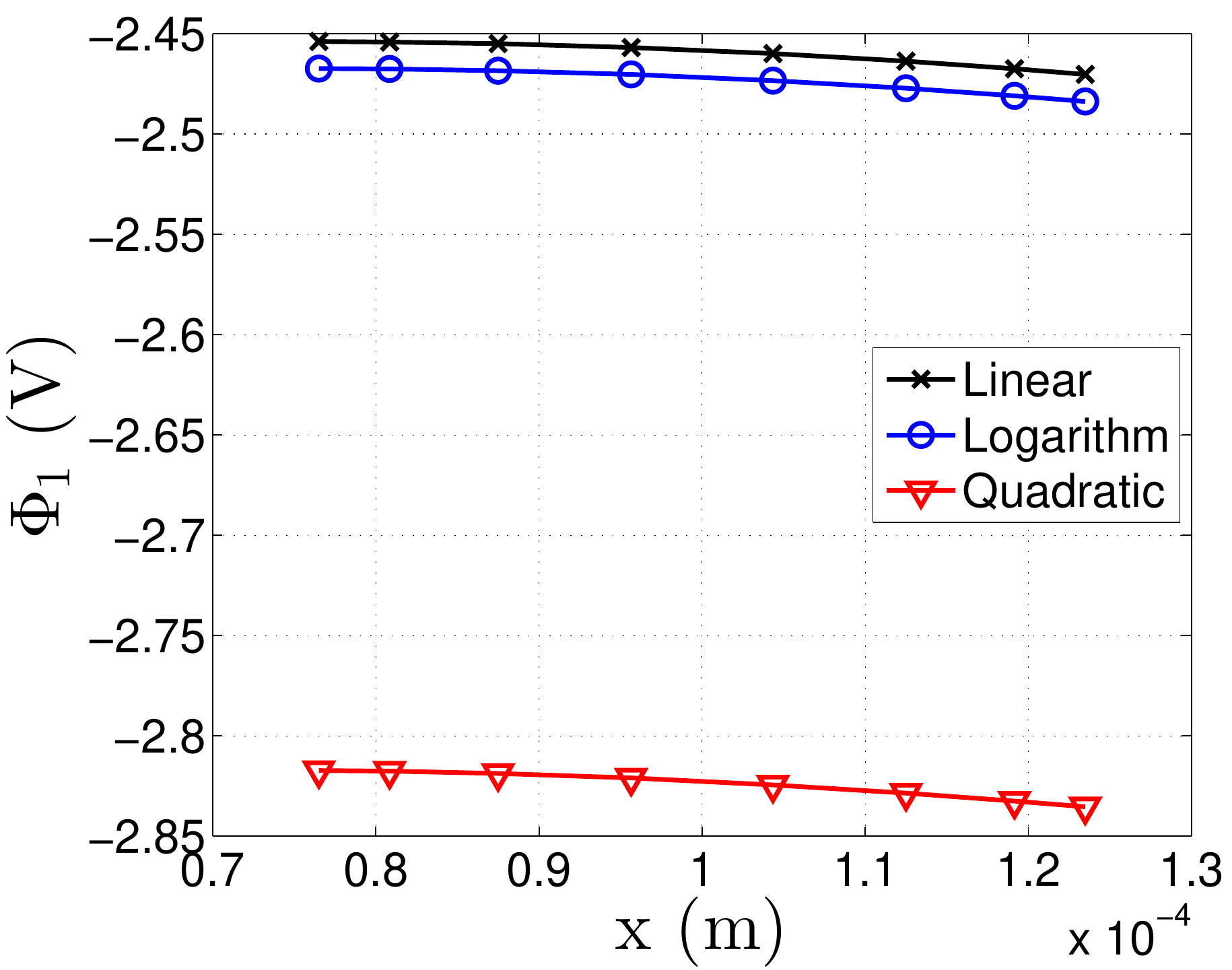}
\caption{Potential in the right electrode.}
\label{fig:states_Dilute_phi1R}
\end{subfigure}
\caption{The state distribution of the non-linear models after 23.2 s  of the standard charging profile with the electrolyte concentration reduced from 930 mol/m$^3$ to 250 mol/m$^3$.}
\label{fig:states_Dilute}
\end{figure}

\begin{figure}
\centering
\graphicspath{ {Figures/} }
\hspace{-1pt}
\begin{subfigure}[b]{0.45\textwidth}
\includegraphics[width=\textwidth]{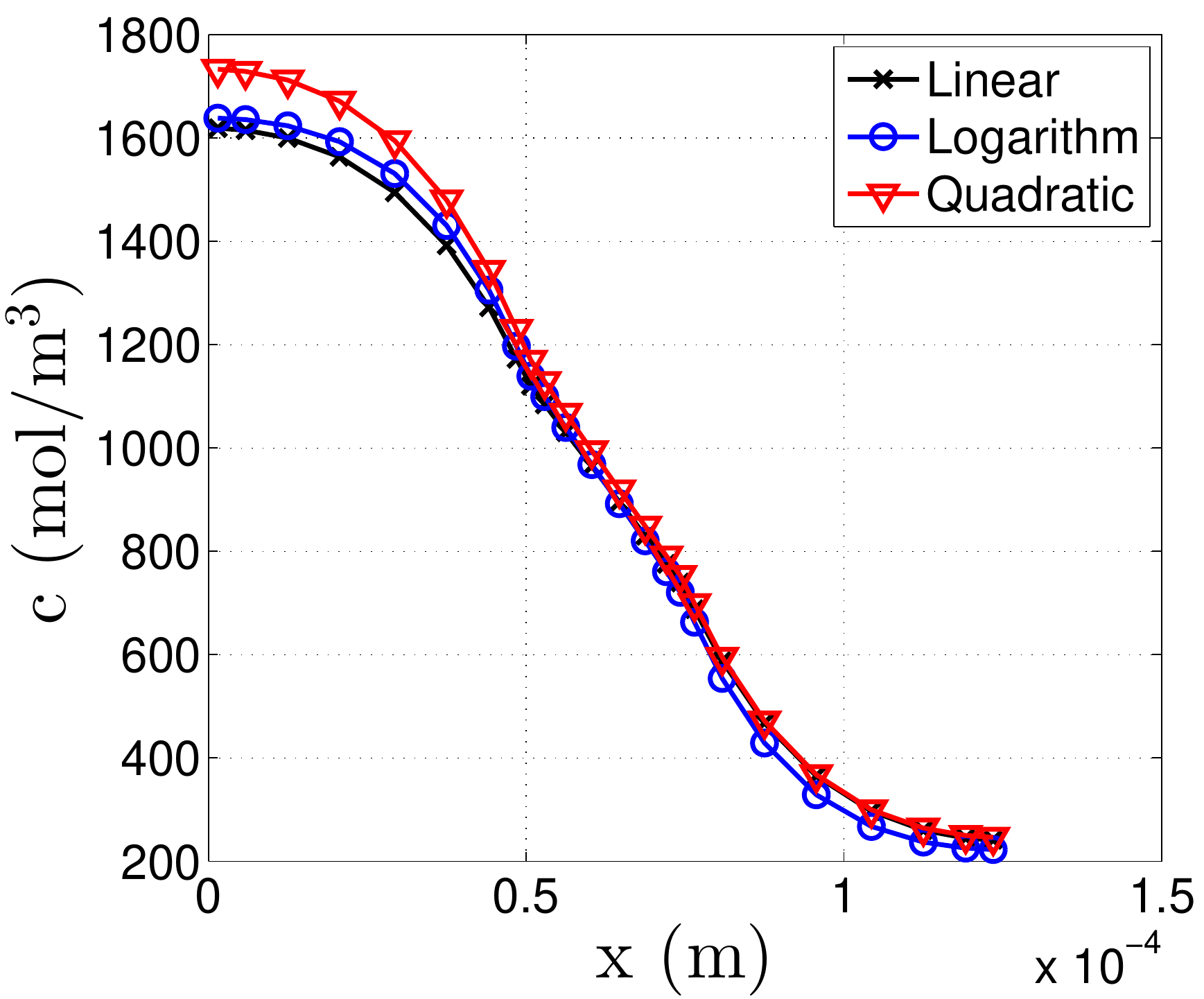}
\caption{Electrolyte concentration.}
\label{fig:states_Extended_c}
\end{subfigure}
\quad
\hspace{-1pt}
\begin{subfigure}[b]{0.475\textwidth}
\includegraphics[width=\textwidth]{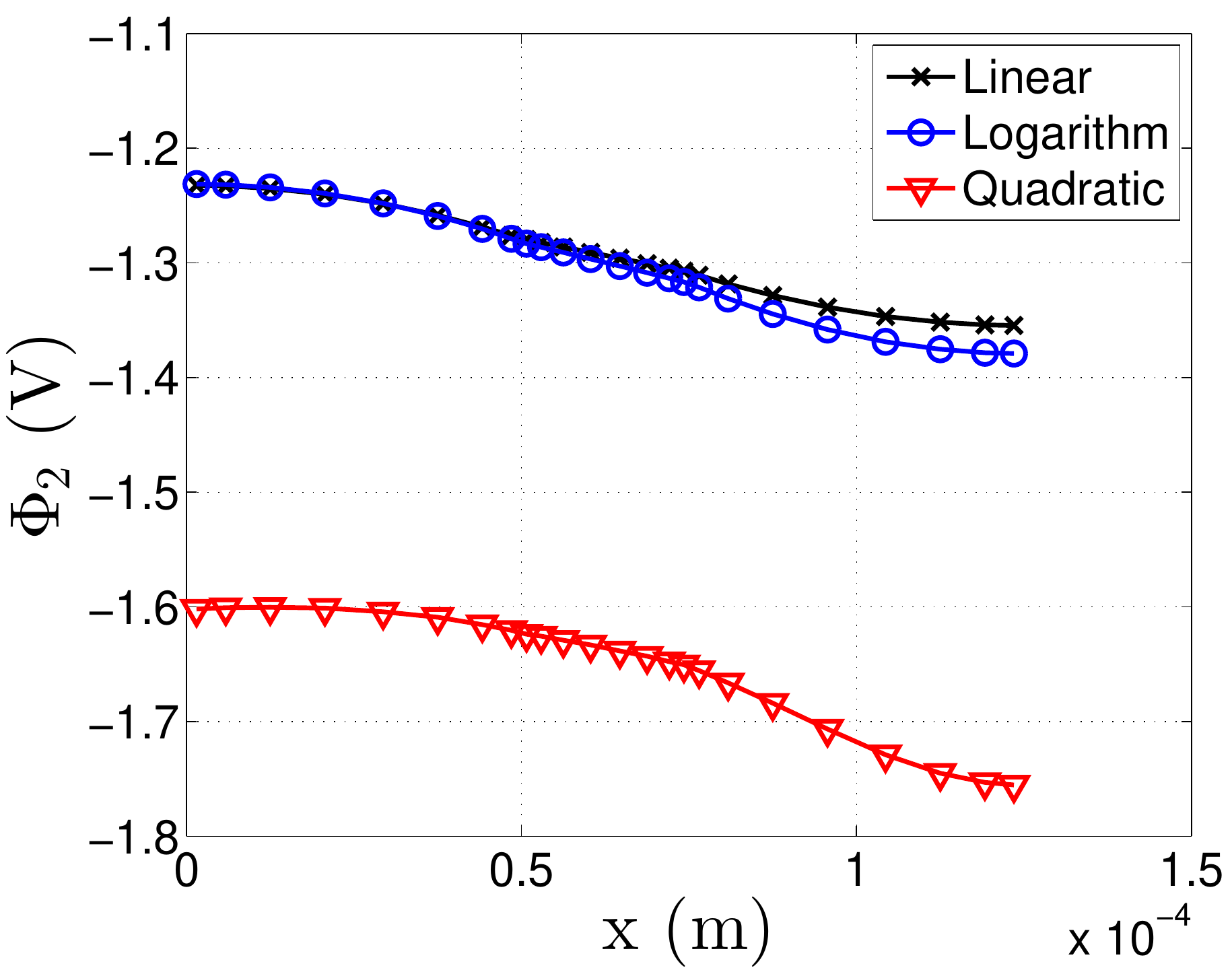}
\caption{Electrolyte potential.}
\label{fig:states_Extended_phi2}
\end{subfigure}
\\
\hspace{2pt}
\hspace{-15pt}
\begin{subfigure}[b]{0.51\textwidth}
\includegraphics[width=\textwidth]{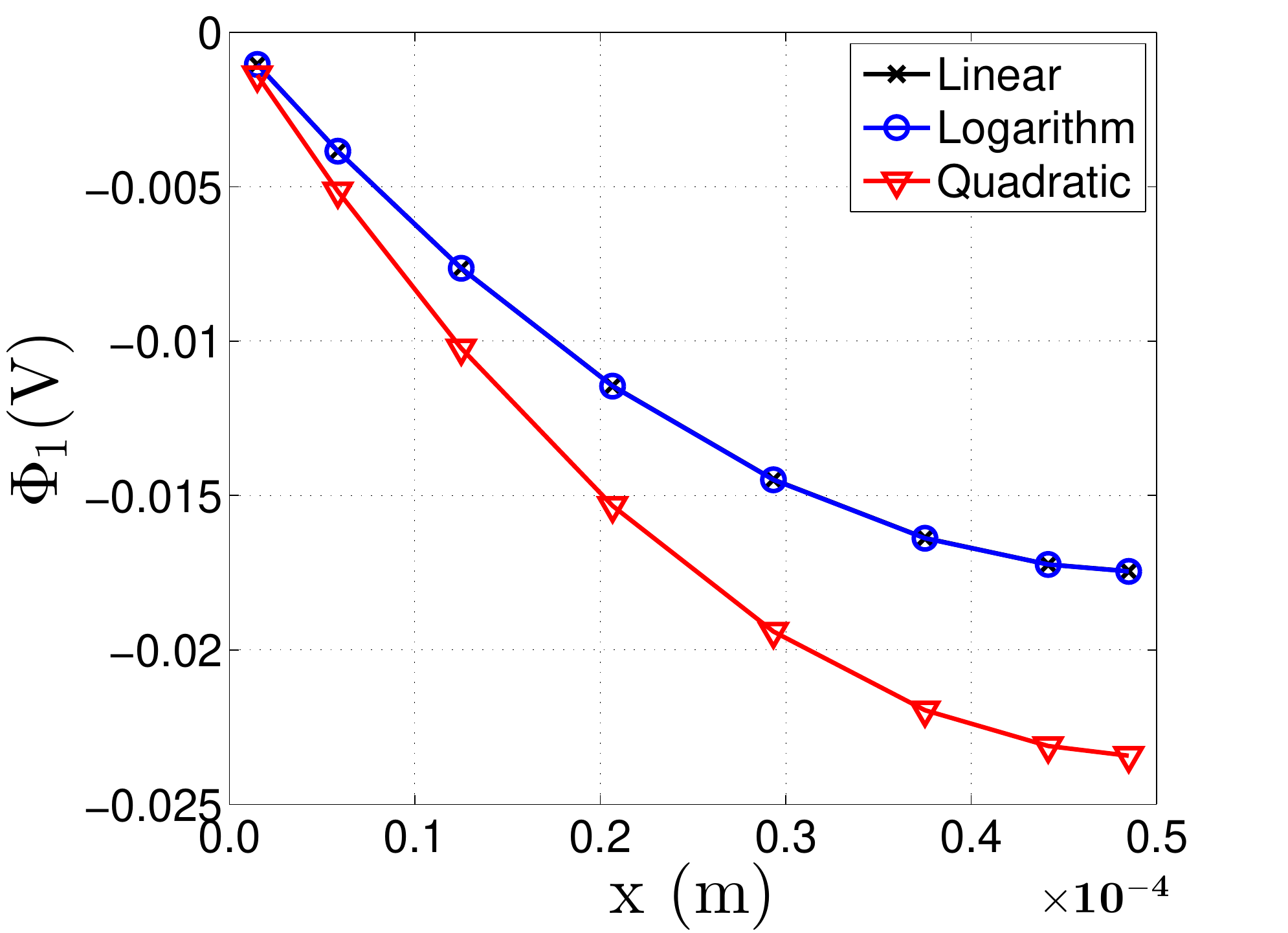}
\caption{Potential in the left electrode.}
\label{fig:states_Extended_phi1L}
\end{subfigure}
\hspace{-6pt}
\begin{subfigure}[b]{0.47\textwidth}
\includegraphics[width=\textwidth]{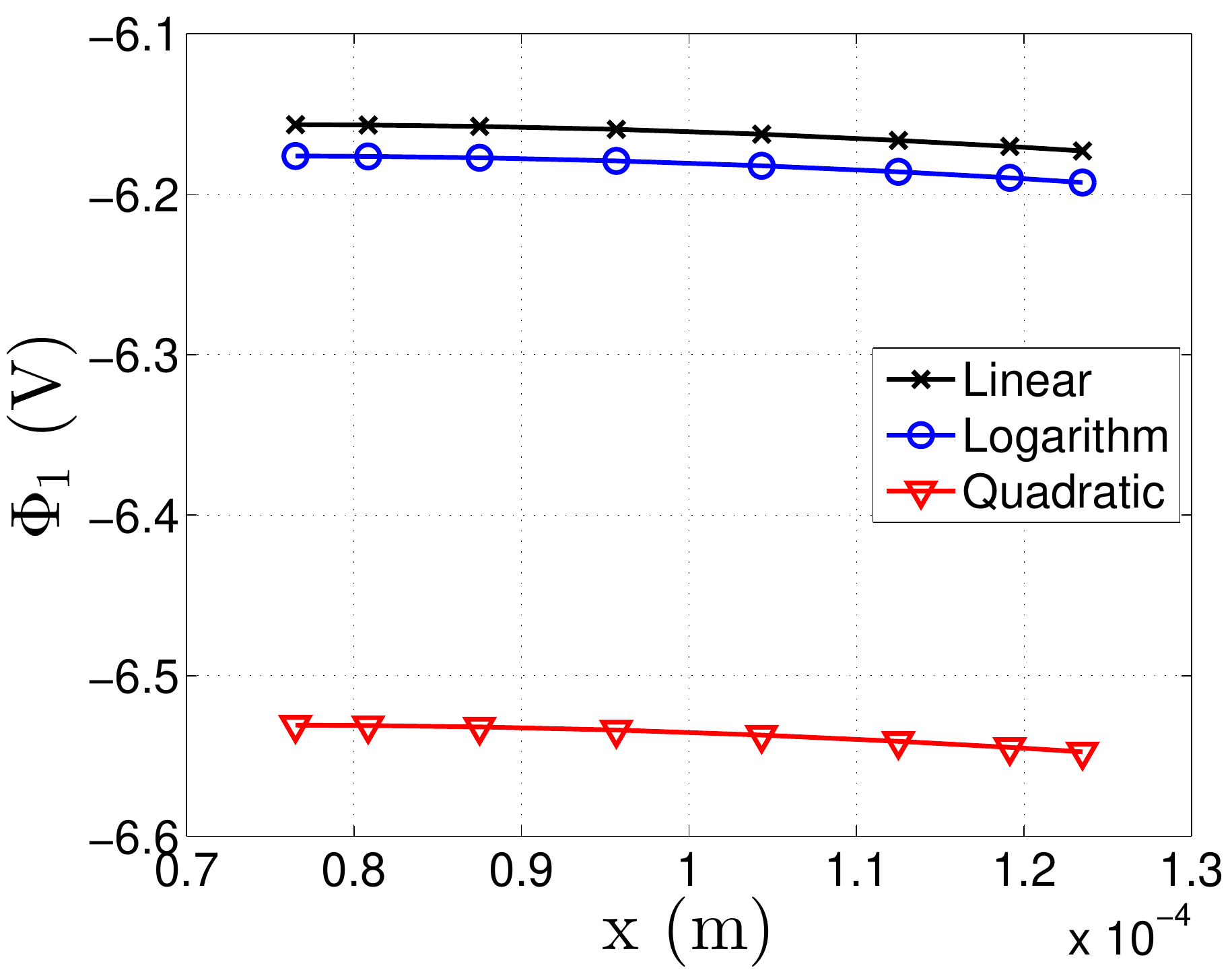}
\caption{Potential in the right electrode.}
\label{fig:phi1R}
\end{subfigure}
\caption{The state distribution of the non-linear models after 130 s  of the extended charging profile.}
\label{fig:states_Extended}
\end{figure}

The effect of the logarithmic and quadratic model non-linearities is investigated in Figures \ref{fig:output_Verb}, \ref{fig:states_Verb}, \ref{fig:states_Dilute} and \ref{fig:states_Extended}. 

It is noted that the choice of $t_+= 0.5$ from Table \ref{tab:GlobalParams} eliminates the non-linear logarithmic term from (\ref{discrete_state_space_3}) and (\ref{eqn:seperator_state_discrete}). To show that the presented model is also applicable for the non-linear case, CC-CV simulations where $t_+$ is increased from 0.5 to 0.75 are run. 
Referring to the labels of Figures \ref{fig:output_Verb}, \ref{fig:states_Verb}, \ref{fig:states_Dilute}
and \ref{fig:states_Extended}, the "Linear" model involves (\ref{discrete_state_space_3}) and (\ref{eqn:seperator_state_discrete}) with $t_+ = 0.5$, the "Logarithm" model involves (\ref{discrete_state_space_3}) and (\ref{eqn:seperator_state_discrete}) with $t_+ = 0.75$ and the "Quadratic" model uses (\ref{electrode_quadratic_discrete}) and (\ref{eqn:seperator_quadratic_discrete}) with $t_+ = 0.5$.
For the standard CC-CV charging profile, the impact of the non-linearities can be seen to be neglible, as shown in Figures \ref{fig:output_Verb} and \ref{fig:states_Verb}, with the numerical solutions being in close agreement with the linear model.
However, in Figure \ref{fig:states_Dilute}, the ionic concentration is decreased from 930 mol/m$^3$ to 250 mol/m$^3$, while in Figure 
\ref{fig:states_Extended}, an extended charging profile is implemented.
This extended charging profile is identical to the standard charging profile except with the initial voltage being -2.37 V, the CC and CV charges being held for 130 s and 70 s respectively and the voltage drop at the CC-CV transition point at 130 s being -1.08 V.
For both of these cases, a noticeable difference in the numerical solutions between the quadratic model and the linear model is observed. 
This implies that the quadratic model is applicable for simulations where there is a large relative change in the electrolyte concentration.
However, for the charging profiles simulated in this paper, the effect of the logarithmic non-linear term is neglible, with there being hardly any difference between its solution and that of the linear model.
Due to the improved convergence properties of spectral methods outlined in Figure \ref{fig:convergence}, all of the non-linear solutions of Figures \ref{fig:output_Verb}, \ref{fig:states_Verb}, \ref{fig:states_Dilute} 
and \ref{fig:states_Extended} are discretised using the SEM. 
This means that the inclusion of the non-linear terms does not lead to a rapid rise in the number of elements, with the solutions of Figure \ref{fig:output_Verb} being obtained using 5 elements in each domain only.

\begin{figure}[h!]
\centering
\graphicspath{ {Figures/} }
\begin{subfigure}[b]{0.5\textwidth}
\includegraphics[width=\textwidth]{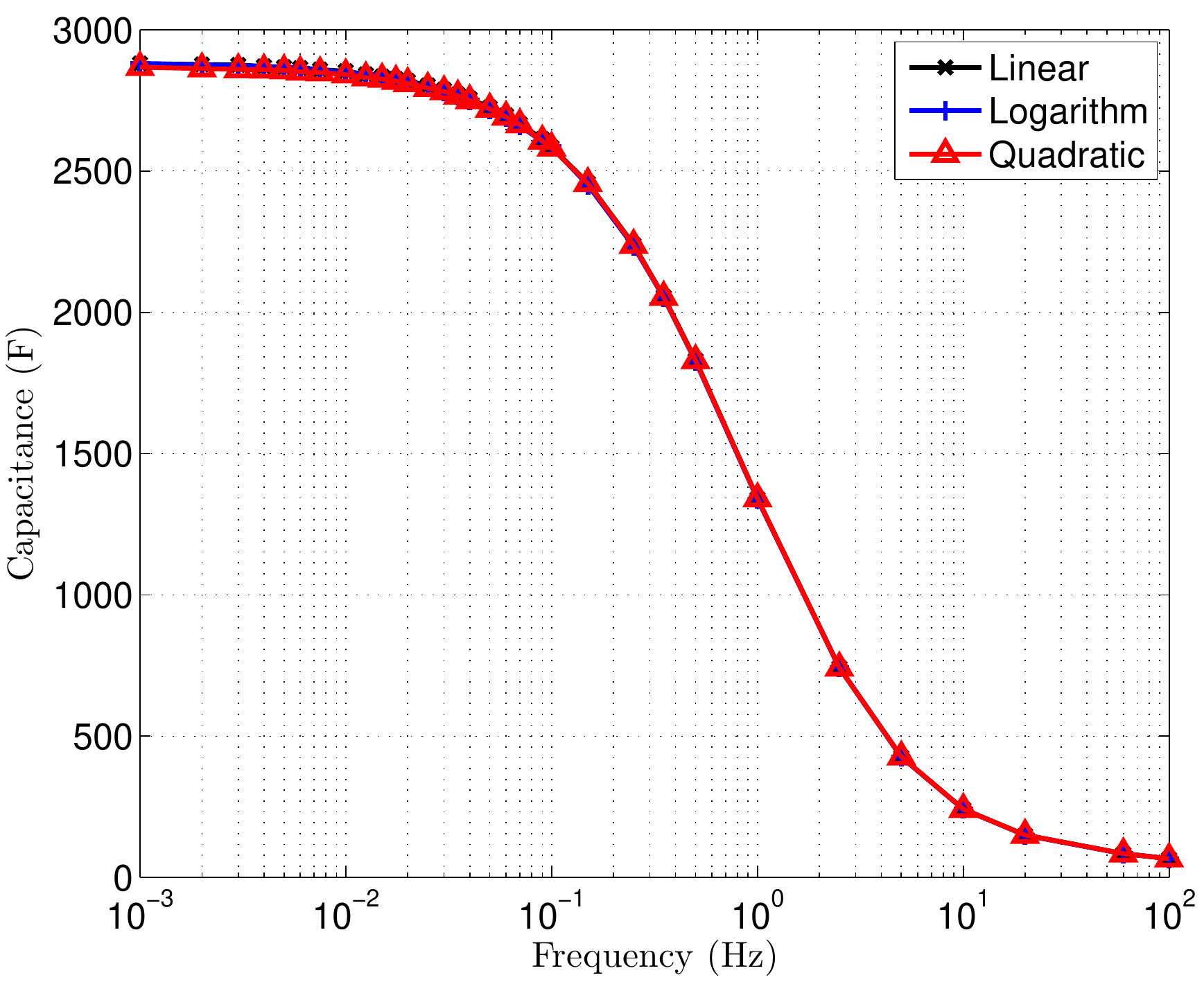}
\caption{Capacitance.}
\label{fig:total_cap}
\end{subfigure}
\\
\hspace{1pt}
\begin{subfigure}[b]{0.492\textwidth}
\includegraphics[width=\textwidth]{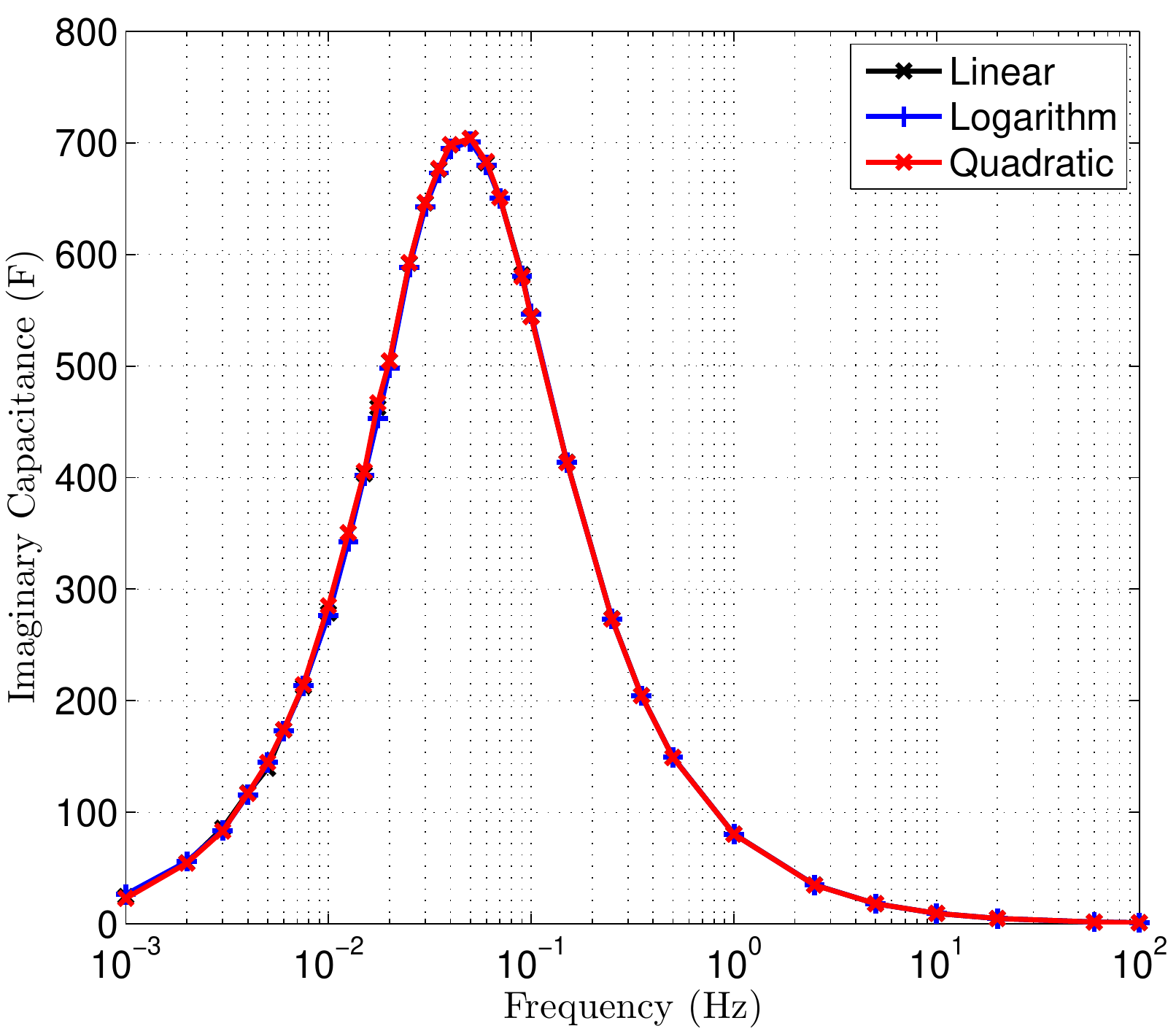}
\caption{Imaginary component of complex capacitance.}
\label{fig:imag_cap}
\end{subfigure}
\caption{The variation of the total capacitance and the imaginary component of the complex capacitance with input current frequency. Obtained using an electrochemical impedance spectroscopy simulation with an input current of 2 A overlaid with a small-amplitude 0.1 A sinusoidal signal.}
\end{figure}

Supercapacitors are often combined with fuel cells or batteries, forming a hybrid power system \cite{Wu20147885}. In this arrangement, the inclusion of the supercapacitor can act as a low-pass filter to reduce the stress on the fuel cell/battery and provide performance benefits, due to the high power density. A key property for the supercapacitor in this application is the variation of capacitance with frequency. Figure \ref{fig:total_cap} shows simulated electrochemical impedance spectroscopy results using the procedure of \cite{lu2013supercapacitors} with an input current of 2 A overlaid with a small-amplitude 0.1 A sinusoidal signal. 
A rapid decrease in capacitance due to diffusion limitations can be observed for all of the models above a certain frequency, known as the knee frequency. This indicates that the performance of the supercapacitor is poor in this operating regime. The knee frequency can be determined by inspecting the peak value of the imaginary component of the complex capacitance \cite{lu2013supercapacitors}, as shown in Figure \ref{fig:imag_cap}. 

In the models used in this study, the capacitance is set as a fixed parameter that is independent of voltage whereas in real supercapacitors, capacitance varies with voltage \cite{Wu20147885}. The model could be extended to incorporate this effect, e.g. by implementing a Guoy-Chapman-Stern capacitance model \cite{lu2013supercapacitors}, however, it is noted that the theoretical understanding of the relationship between double layer potential and capacitance is as yet too complex \cite{lu2013supercapacitors}.

\section{Conclusion}
In this paper, two non-linear physics-based supercapacitor models were implemented in a novel way using a spectral element method. The first model was based on \cite{verbrugge2005microstructural} while the second model accounted for the linear dependence between electrolyte conductivity and concentration. For a typical supercapacitor CC-CV charging profile, the numerical solutions from both models were found to be in close agreement with experimental data.
However, for other conditions, such as the electrolyte concentration diluted or the charge duration of the supercapacitor extended, a noticeable difference between the numerical solutions of the two models was observed. 
This was due to the large relative change in electrolyte concentration affecting the conductivity. 
Electrical impedance spectroscopy simulations were also carried out on the models, and it was found that the capacitance of the supercapacitor models decreased rapidly when the frequency of the input current exceeded an upper threshold, as expected.

The normalised error of the model numerical solutions discretised using the spectral element method was found to converge faster as the number of domain elements was increased compared to the finite difference method of \cite{verbrugge1994finite}. 
An accurate solution could also be obtained using fewer elements than the finite element 3D model of \cite{allu2014generalized}.
As such, discretising the models using the spectral element method reduced the number of nodes needed to obtain a specified solution tolerance and resulted in a lower order model that was faster to implement. This implies that the models could be appropriate for a real-time implementation as well as for accurate state estimation with an observer.


\section*{Acknowledgments}
Support from the UK Engineering and Physical Sciences Research Council is gratefully acknowledged.

\section*{References}
\bibliography{bibliog} 

\newpage
\begin{table} 
\caption*{Nomenclature}
\label{Symbols}
\centering 
\renewcommand{\arraystretch}{1.3}
\begin{tabular}{|c| c| c|}
\hline 
Symbol &Definition & Units \\
\hline
$c$ & Electrolyte concentration & $\text{mol}/\text{m}^3$ \\
$\Phi_1$ & Electrode potential & V \\
$\Phi_2$ & Electrolyte potential & V \\
$I$ & Current & A \\
$S$ & Electrode surface area & m$^2$ \\
$i$ & Current density& A/m$^2$ \\
$i_1$ & Current density in the solid phase & A/m$^2$ \\
$i_2$ & Current density in the liquid phase & A/m$^2$ \\
$L_{elect}$ & Electrode length & m \\
$L_{sep}$ & Separator length & m \\
$N$ &Number of elements in the domain & \\
$aC$ &Specific capacitance & F/m$^3$ \\
$U$ &Ionic flux & mol/m$^2$s\\
$R$ & Universal gas constant & J/K mol \\
$F$ & Faraday constant & C mol$^{-1}$\\
$T$ & Temperature & K \\
$\zeta$ & Ionic charge & C \\
$c_0$ & Initial Salt concentration & mol/m$^3$ \\
$\kappa$ & Electrolytic conductivity& S/m \\
$\kappa_{\infty}$ & Free solution electrolytic conductivity & S/m \\
$\sigma$ & Electrode conductivity & S/m \\
$D$ &Diffusion coefficient & m$^2$/s\\
$\Gamma$ &Tortuosity & \\
$t_{j}$ & Ion transference number & \\
$\frac{dq_{j}}{dq}$ & Change in surface concentration of species $j$ associated with & \\
& a Change on the surface charge of electrode & \\
$\epsilon$ & Porosity void fraction & \\
\hline 
\end{tabular}
\label{Symbols}
\end{table}

\end{document}